\documentclass[12pt]{article}
\pdfoutput=1
\usepackage{epsf,amsfonts,amssymb,epsfig,amsmath,graphics,slashed}
\usepackage{hyperref,graphicx, subfig,hep}

\newcommand{\nn}{\notag \\}
\newcommand{\vol}{\mathrm{vol}}

\begin{document}

\makeatletter
\renewcommand{\theequation}{\thesection.\arabic{equation}}
\@addtoreset{equation}{section}
\makeatother

\baselineskip 18pt

\begin{titlepage}

\vfill

\begin{flushright}
Imperial/TP/2014/JG/02\\
\end{flushright}

\vfill

\begin{center}
   \baselineskip=16pt
   {\Large\bf  Flowing from $AdS_5$ to $AdS_3$ with $T^{1,1}$}
  \vskip 1.5cm
     Aristomenis Donos$^1$ and Jerome P. Gauntlett$^2$\\
   \vskip .6cm
    \vskip .6cm
      \begin{small}
      \textit{$^1$DAMTP, 
       University of Cambridge\\ Cambridge, CB3 0WA, U.K.}
        \end{small}\\
      \vskip .6cm
      \begin{small}
      \textit{$^2$Blackett Laboratory, 
        Imperial College\\ London, SW7 2AZ, U.K.}
        \end{small}\\*[.6cm]

\end{center}

\vfill

\begin{center}
\textbf{Abstract}
\end{center}

\begin{quote}
We construct supersymmetric domain wall solutions of type IIB supergravity that interpolate between 
$AdS_5\times T^{1,1}$ in the UV and $AdS_3\times\mathbb{R}^2\times S^2\times S^3$
solutions in the IR. The $\mathbb{R}^2$ factor can be replaced 
with a two-torus and then the solution describes a supersymmetric flow across dimensions, similar to wrapped brane solutions. 
While the domain wall 
solutions preserve $(0,2)$ supersymmetry, the $AdS_3$ solutions in the IR have an enhanced
$(4,2)$ superconformal supersymmetry 
and are related by two T-dualities to the $AdS_3\times S^3\times S^3\times S^1$ type IIB solutions which preserve a large $(4,4)$ superconformal 
supersymmetry. The domain wall solutions exist within the $N=4$ $D=5$ gauged supergravity theory
that is obtained from a consistent Kaluza-Klein truncation of type IIB supergravity on $T^{1,1}$; a feature driving
the flows is that two $D=5$ axion like fields, residing in the $N=4$ Betti multiplet, depend linearly on the two legs of the $\mathbb{R}^2$ factor.

\end{quote}

\vfill

\end{titlepage}
\setcounter{equation}{0}



\section{Introduction}

Type II string theory possesses 
$AdS_3\times S^3\times S^3\times S^1$ solutions which are supported by magnetic three-form 
fluxes threading the $S^3$ factors
as well as electric three-form flux threading the $S^3\times S^3\times S^1$ factor \cite{Cowdall:1998bu,Boonstra:1998yu,Gauntlett:1998kc,Elitzur:1998mm,deBoer:1999rh}. 
While these solutions have been known for a long time the dual field theory, which preserves a large $(4,4)$ super conformal symmetry, remains elusive. A detailed discussion of some of the issues is presented in \cite{Gukov:2004ym} and we note that a recent proposal for the dual field theory 
appears in \cite{Tong:2014yna}. In this paper we will discuss some new results
on these type II solutions, using a rather indirect approach.

Starting with the $AdS_3\times S^3\times S^3\times S^1$ solutions of type IIB string theory with the magnetic and electric three-form
fluxes in the RR sector, we can carry out two T-dualities on two circles to obtain other type IIB solutions with an
$AdS_3$ factor.
If we choose one of the two circles to be the explicit $S^1$ factor and the other to be a diagonal of the two
Hopf fibres of the $S^3\times S^3$, then we obtain the $AdS_3\times T^2\times S^2\times S^3$
solutions with non-trivial RR five-form and three-form fluxes as well as NS three-form flux that were first found in \cite{Donos:2008ug}.
Here we will show that, as solutions of type IIB supergravity, these solutions preserve $(4,2)$ superconformal
symmetry and not just the $(0,2)$ superconformal symmetry that was guaranteed from the original construction of \cite{Donos:2008ug}.

A principal result of this paper is the construction of type IIB supergravity domain-wall solutions that interpolate between
the $AdS_5\times T^{1,1}$ solution in the UV and approach these $AdS_3\times T^2\times S^2\times S^3$ solutions
in the IR. The flow solutions preserve $(0,2)$ Poincar\'e symmetry which is enhanced to
$(4,2)$ superconformal symmetry at the IR fixed point.
The supergravity solutions are constructed directly in type IIB supergravity. However, they can also
be constructed in an $N=4$ $D=5$ gauged supergravity theory that can be obtained as a consistent KK truncation
of type IIB supergravity on $T^{1,1}$ \cite{Cassani:2010na,Bena:2010pr,Liu:2011dw} (extending
\cite{Cassani:2010uw,Liu:2010sa,Gauntlett:2010vu,Skenderis:2010vz,Bah:2010cu,Liu:2010pq}). 
This perspective is helpful in identifying the deformations of the $N=1$ SCFT that are needed to flow to the $AdS_3$ fixed points.

As is well known, the $AdS_5\times T^{1,1}$ 
UV fixed point is dual to an $N=1$ SCFT in $D=4$ that arises on D3-branes
sitting at the apex of the conifold \cite{Klebanov:1998hh}. 
Our constructions can be viewed as a variation of wrapped-brane solutions 
\cite{Maldacena:2000mw} (see \cite{Gauntlett:2003di} for a review and \cite{Benini:2012cz,Benini:2013cda,Bobev:2014jva} for recent constructions with $AdS_3$ factors), with
the D3-branes wrapping a $T^2$ and sitting at the apex of the conifold with particular deformations
switched on. In particular, an important
ingredient is that there are two axion like fields in the Betti multiplet of the gauged-supergravity
which are linear in the $T^2$ directions. This mechanism for preservation of supersymmetry differs
from the usual one of activating $R$-symmetry currents, related to the spin connection of the cycle being wrapped,
and also the constructions of \cite{Almuhairi:2011ws,Donos:2011pn} where there are magnetic fluxes threading a $T^2$. In the most general solutions that we construct here, though,
there is also a magnetic flux of the Betti vector field threading the $T^2$ factor.

We also analyse the flux-quantisation for the $AdS_3\times T^2\times S^2\times S^3$ fixed point solutions.
This turns out to be somewhat subtle due to the presence of Page charges. While the quantisation of Page charges
have been discussed before \cite{Marolf:2000cb,Aharony:2011yc,Assel:2011xz,deBoer:2012ma}, analysing our solutions
reveals some new issues, which will also arise in the context other classes of solutions. 
We explain our prescription for quantising the Page charges
and use this to obtain the central charge of the dual SCFT. 

The above discussion focussed on solutions that flow from $AdS_5\times T^{1,1}$ to
$AdS_3\times T^2\times S^2\times S^3$. However, if one does not compactify two spatial dimensions
then one has solutions flowing from 
$AdS_5\times T^{1,1}$ to $AdS_3\times \mathbb{R}^2\times S^2\times S^3$.
Such solutions may have interesting applications in the context of applied AdS/CFT, where there has been various studies
on the emergence of $AdS_3$ solutions after switching on magnetic fields, including examples preserving supersymmetry
\cite{Almuhairi:2011ws,Donos:2011pn}. 
Our solutions, utilising axions, provide 
an alternative approach\footnote{Axions/massless fields that are linear in either null or spatial coordinates
have been used in applied AdS/CFT in other works including \cite{Balasubramanian:2010uk,Donos:2010tu,Azeyanagi:2009pr,Mateos:2011ix,Donos:2013eha,Andrade:2013gsa}.} 
to hitting such fixed points.
It is also worth commenting that our type IIB domain wall solutions share some similarities with supersymmetric solutions of $D=11$ supergravity that 
interpolate between $AdS_4\times Q^{111}$
in the UV and supersymmetric 
$AdS_2\times\mathbb{R}^2\times S^2\times S^2\times S^2\times S^1$ solutions in the IR
\cite{Donos:2012sy}; a difference, however, is that those flows were driven
by electric and magnetic baryonic fluxes.

The plan of the rest of the paper is as follows. In section 2 we describe the supersymmetric domain wall solutions
in the simplest setting and then generalise them to a one-parameter family of flows in section 3. 
We briefly conclude in section 4.
We have three appendices. In appendix A we review the charge quantisation of the $AdS_3\times S^3\times S^3\times S^1$ solutions. In appendix B we demonstrate that the fixed point solutions preserve $(4,2)$
supersymmetry and in appendix C we discuss some additional aspects of the quantisation of Page charges.

\section{A flow from $AdS_{5}\times T^{1,1}$ to $AdS_{3}\times\mathbb{R}^2\times S^2\times S^3$}\label{basicflow}

\subsection{General set-up}
We will construct supersymmetric solutions of type IIB supergravity \cite{Schwarz:1983qr,Howe:1983sra} using the conventions 
given in \cite{Gauntlett:2005ww}. We will consider solutions with trivial axion and dilaton and hence the R-R and
NS-NS three-forms can be combined into the complex three-form $G=-dB-idC$, where $B,C$ are both two-forms. 
The Bianchi identities
for $G$ and the self-dual five-form $F$, satisfying $F=*F$, are given by
\begin{align}\label{bianch}
dG=0,\qquad dF=\frac{i}{2}G\wedge G^*\,,
\end{align}
while the equation of motion for $G$ can be written
\begin{align}\label{Geom}
\nabla^\mu G_{\mu\nu\rho}=-\frac{i}{6}F_{\nu\rho\sigma_1\sigma_2\sigma_3}G^{\sigma_1\sigma_2\sigma_3}\,.
\end{align}
The Killing spinor equations take the form
\begin{align}
\nabla_{\mu}\varepsilon+\frac{i}{16}\,\slashed F \Gamma_{\mu}\varepsilon+\frac{1}{16}\,\left(\Gamma_{\mu}\slashed G+2\slashed G \Gamma_{\mu}\right)\varepsilon^{c}=&0\,,\label{eq:kill1}\\
\slashed G\varepsilon=&0\,.\label{eq:kill2}
\end{align}

We begin by recalling the standard $AdS_5\times T^{1,1}$ solution \cite{Romans:1984an} of type IIB supergravity.
The metric and the self-dual five-form are given by
\begin{align}\label{t11sol}
\frac{1}{L^2}ds^{2}=& e^{2\rho}\,\left(-dt^{2}+dx^{2}+dx_{1}^{2}+dx_{2}^{2}\right)+d\rho^{2}+\frac{1}{6}\left(ds_{1}^{2}+ds_{2}^{2} \right)+\eta^{2}\,,\notag\\
\frac{1}{L^4}F=&4e^{4\rho}\,dt\wedge dx\wedge dx_{1}\wedge dx_{2}\wedge d\rho+\frac{1}{9}\eta\wedge\mathrm{vol}_{1}\wedge \mathrm{vol}_{2}\,,
\end{align}
where we have defined 
\begin{align}\label{adsf}
ds_{i}^{2}&=\left(d\theta_{i}^{2}+\sin^{2}\theta_{i}\,d\phi_{i}^{2} \right),\qquad
\mathrm{vol}_i=\sin\theta_{i}\,d\theta_{i}\wedge d\phi_{i}\,,\notag\\
\eta&=\frac{1}{3}\left(d\psi+P\right),\qquad P=P_1+P_2,\qquad P_i=-\cos\theta_i  d\phi_i\,,
\end{align}
and $dP_i=\mathrm{vol}_i$.
Note that $\eta$ is the Reeb one-form, $\partial_\psi$ is the Reeb Killing vector and the period of $\psi$ is $4\pi$.
Also, $L$ is a constant length scale fixed by flux quantisation (given in \eqref{Nval} below).
This solution preserves four Poincar\'e and four superconformal supersymmetries. It is useful to record the explicit form of
the Poincar\'e supersymmetries. Using the obvious orthonormal frame (see \eqref{ortho} below)
the Poincar\'e supersymmetries
satisfy the following algebraic conditions\footnote{Note that our conventions are such that $\Gamma_{0123456789}\varepsilon=-\varepsilon$ and also $\varepsilon_{0123456789}=+1$.}
\begin{align}\label{eq:projections}
&i\Gamma^{0123}\varepsilon=-\varepsilon\,,\nn
&\Gamma^{56}\varepsilon=i\varepsilon,\qquad \Gamma^{78}\varepsilon=i\varepsilon,\qquad \Gamma^{49}\varepsilon=i\varepsilon\,,
\end{align}
These conditions are equivalent to $\Gamma^{5678}\varepsilon=-\varepsilon$, $\Gamma^{5649}\varepsilon=-\varepsilon$, corresponding
to the conifold (the Calabi-Yau cone over $T^{1,1}$), combined with $i\Gamma^{0123}\varepsilon=-\varepsilon$ corresponding
to putting a D3-brane at its apex.
The four Poincar\'e Killing
spinors can be written $\varepsilon=e^{\rho}\varepsilon_0$ where $\varepsilon_0$
satisfies 
\begin{align}\label{second}
\hat{\nabla}_{m}\varepsilon_0-\frac{1}{2}\Gamma^{4}\Gamma_{m}\varepsilon_0=0\,,
\end{align}
where $\hat\nabla$ is the Levi-Civita connection on $T^{1,1}$ with coordinates $y^m$.

We are interested in constructing supersymmetric domain walls that approach $AdS_5\times T^{1,1}$ in the UV, and
flow to particular $AdS_3\times M_7$ solutions in the IR. The ansatz that we shall consider first is given by
\begin{align}\label{ansatz}
\frac{1}{L^2}ds^{2}&= e^{2A}\,\left(-dt^{2}+dx^{2}\right)+e^{2B}\,\left(dx_{1}^{2}+dx_{2}^{2}\right)+d\rho^{2}+\frac{1}{6}e^{2U}\,\left(ds_{1}^{2}+ds_{2}^{2} \right)+e^{2V}\,\eta^{2}\,,\notag\\
\frac{1}{L^4}F&=4e^{2A+2B-V-4U}\,dt\wedge dx\wedge dx_{1}\wedge dx_{2}\wedge d\rho+\frac{1}{9}\,\eta\wedge\mathrm{vol}_{1}\wedge \mathrm{vol}_{2}\,,\notag\\
&+\frac{\lambda^2}{12}\left[\,dx_{1}\wedge dx_{2}\wedge \eta\wedge \left(\mathrm{vol}_{1}+\mathrm{vol}_{2}\right)+e^{2A-2B-V}\,dt\wedge dx\wedge d\rho\wedge\left(\mathrm{vol}_{1}+\mathrm{vol}_{2}\right)\right]\,,\notag\\
\frac{1}{L^2}G=&\frac{\lambda}{6}\,(dx_{1}-i dx_{2})\wedge\,\left(\mathrm{vol}_{1}-\mathrm{vol}_{2} \right)\,,
\end{align}
where $\lambda$ is a constant and 
$A,B,U,V$ are functions of $\rho$ only. We will discuss the dual SCFT interpretation of this ansatz in section \ref{scftint} and discuss a
generalisation in section \ref{genflow}.

Observe that, by construction, the ansatz has a self-dual five-form, $F=*F$,
and that both the Bianchi identities \eqref{bianch} and the equation of motion for $G$ 
\eqref{Geom} are satisfied.
To analyse the conditions for preservation of supersymmetry, \eqref{eq:kill1} and \eqref{eq:kill2}, we use the orthonormal frame
\begin{align}\label{ortho}
e^{0}&=e^{A}\,dt,\quad e^{1}=e^{A}\,dx,\quad e^{2}=e^{B}\,dx_{1},\quad e^{3}=e^{B}\,dx_{2},\quad e^{4}=d\rho,\notag\\
e^{5}&=\frac{e^{U}}{\sqrt 6}d\theta_1,\quad e^{6}=\frac{e^{U}}{\sqrt 6}\sin\theta_1d\phi_1,\quad e^{7}=\frac{e^{U}}{\sqrt 6}\,d\theta_{2},\quad e^{8}=\frac{e^{U}}{\sqrt 6}\sin\theta_2 d\phi_2,\quad e^{9}=e^{V}\eta\,.
\end{align}
We will continue to impose the algebraic conditions \eqref{eq:projections} and we will also impose $\Gamma^{23}\varepsilon=i\varepsilon$ or equivalently
\begin{align}\label{proj2}
\Gamma^{01}\varepsilon=\varepsilon\,,
\end{align}
corresponding to a chiral $(0,2)$ Poincar\'e supersymmetry in $d=1+1$.
It is straightforward to see that
\eqref{eq:kill2} is automatically satisfied while equation \eqref{eq:kill1} reduces to
\begin{equation}
\nabla_{\mu}\varepsilon+\frac{i}{16}\,\slashed F \Gamma_{\mu}\varepsilon=0\,.
\label{eq:kill3}
\end{equation}
A calculation now shows that we can solve \eqref{eq:kill3} provided that we choose $\varepsilon=e^{A/2}\varepsilon_0$ with $\varepsilon_0$
satisfying \eqref{second}, and that the functions $A,B,U,V$ satisfy the following coupled first order differential
equations
\begin{align}\label{eq:first_order_equations}
A^{\prime}-e^{-V-4U}-\frac{\lambda^{2}}{4}e^{-2B-V-2U}&=0\,,\notag\\
B^{\prime}-e^{-V-4U}+\frac{\lambda^{2}}{4}e^{-2B-V-2U}&=0\,,\notag\\
U^{\prime}+e^{-V-4U}-e^{V-2U}&=0\,,\nn
V^{\prime}-3\,e^{-V}+2\,e^{V-2U}+e^{-V-4U}+\frac{\lambda^{2}}{4}e^{-2B-V-2U}&=0\,.
\end{align}
Since our ansatz satisfies $F=*F$, the Bianchi identities \eqref{bianch} and the equation of motion for $G$ \eqref{Geom}, we can conclude from the result in
appendix D of \cite{Gauntlett:2005ww}, that
any solution to these differential equations will also solve the type IIB Einstein equations and hence
gives rise to a supersymmetric solution of type IIB supergravity preserving at least
two supersymmetries.

We immediately recover the $AdS_5\times T^{1,1}$ solution \eqref{adsf} by setting $\lambda=0$ and
\begin{align}\label{ads5fp}
A=B=\rho,\quad U=V=0.\qquad
\end{align}
When $\lambda\ne 0$, it is convenient to scale the coordinates $x_i$ and shift the function $B$ by a constant to set $\lambda=2$, without loss of generality.
We then find another exact solution to \eqref{eq:first_order_equations} corresponding to a solution with an $AdS_3$ factor:
\begin{equation}\label{ads3fp}
A=\frac{3^{3/4}}{\sqrt{2}}\rho,\quad B=\frac{1}{4}\ln\left(\frac{4}{3}\right),\quad U=\frac{1}{4}\ln\left(\frac{4}{3}\right),\quad V=-\frac{1}{4}\ln\left(\frac{4}{3}\right)\,.
\end{equation}
Indeed, if we substitute this solution into the ansatz \eqref{ansatz} and
scale $x_{i}= \frac{1}{6^{1/2}} z_{i}$ we find that the metric can be written as
\begin{align}\label{irfixed}
\frac{1}{L^2}ds^{2}&=\frac{1}{3^{3/2}}\,\left(2\,ds^{2}\left(AdS_{3}\right)+dz_{1}^{2}+dz_{2}^{2}+d{s}^{2}_{1}+d{s}_{2}^{2}+\frac{1}{2}\,\left(d\psi+P \right)^{2} \right)\,,\nn
\frac{1}{L^4}F&=\frac{1}{27}\Bigg(  \mathrm{vol}(AdS_3)\wedge\left[4dz_1\wedge dz_2+2 (\mathrm{vol}_1+\mathrm{vol}_2)\right]  \nn
&\qquad\qquad+\left(d\psi+P \right)\wedge\left[\mathrm{vol}_1 \wedge \mathrm{vol}_2+\frac{1}{2} dz_1\wedge dz_2\wedge (\mathrm{vol}_1+\mathrm{vol}_2) \right]\Bigg)\,,\nn
\frac{1}{L^2}G&=\frac{1}{3^{3/2}2^{1/2}}(dz_1-i dz_2)\wedge\left( \mathrm{vol}_1-\mathrm{vol}_2\right)\,.
\end{align}
This solution was first found\footnote{To compare we should set, in the notation of \cite{Donos:2008ug}, $l_{1}=l_{2}=1$, $m_{1}=1/2$ and also identify $\psi=z$ and 
$L^2=(3^{3/2}/2^{1/2})L^2_{there}$.} in section 3.1.2 of \cite{Donos:2008ug}.  
Observe that the topology of the internal five-dimensional compact space is unchanged from that of $T^{1,1}$, namely $S^2\times S^3$.
Thus the topology of the $D=10$ solution is $AdS_3\times \mathbb{R}^2\times S^2\times S^3$, or, if we take
$x_i$ (or equivalently the $z_i$) to parametrise a two-torus, $AdS_3\times T^2\times S^2\times S^3$.

By construction this $AdS_3$ solution preserves $(0,2)$ Poincar\'e supersymmetry and this is supplemented by a further two supersymmetries to give $(0,2)$ superconformal symmetry. In fact this was already known from
the construction in \cite{Donos:2008ug}.
However, as we show in appendix \ref{appes}, and further discuss in section \ref{enhsus}, the fixed point
actually preserves an enhanced $(4,2)$ superconformal supersymmetry (i.e. twelve supersymmetries in total).

\subsection{The supersymmetric flow}

We would now like to construct, numerically, a supersymmetric flow from the $AdS_5\times T^{1,1}$ solution
to the $AdS_3\times \mathbb{R}^2 \times S^2\times S^3$ solution \eqref{irfixed}. We will develop a series
expansion of the differential equations \eqref{eq:first_order_equations}
about both the $AdS_5$ UV fixed point \eqref{ads5fp} and the $AdS_3$ IR fixed point \eqref{ads3fp} and then use a shooting technique
to match them. We again set $\lambda=2$.

By expanding about the $AdS_5$ UV fixed point \eqref{ads5fp} we can develop the following expansion as $\rho\to\infty$:
\begin{align}\label{uvexp}
A&=
\rho-\frac{5 }{12}e^{-2 \rho}+\frac{287 }{1152}e^{-4 \rho}-\frac{5953 }{34560}e^{-6 \rho}+\dots\,,\nn
B&=\rho+\frac{7 }{12}e^{-2 \rho}-\frac{385 }{1152}e^{-4 \rho}+\frac{8267 }{34560}e^{-6 \rho}+\dots\,,\nn
U&=\frac{1}{12}e^{-2 \rho}-\frac{13 }{96}e^{-4 \rho}+{c_1} e^{-6 \rho}+\frac{3}{20} e^{-6 \rho} \rho+\dots\,,\nn
V&=-\frac{1}{6}e^{-2 \rho}+\frac{37 }{96}e^{-4 \rho}+\left(\frac{9023}{51840}-4 {c_1}\right) e^{-6 \rho}-\frac{3}{5} e^{-6 \rho} \rho+\dots\,.
\end{align}
Here we have used the freedom to shift $A$ by a constant in \eqref{eq:first_order_equations} to eliminate an integration constant.
Notice that the expansion depends on one constant $c_1$. We will comment on the dual $D=4$ SCFT interpretation of this UV expansion in the next subsection.

We now consider the expansion about the $AdS_3$ fixed point \eqref{ads3fp}. We find that as $\rho\to-\infty$ it is fixed by three integration constants, $a_0,s_1$ and $s_2$:
\begin{align}\label{irexp1}
A&=a_{0}+\rho/R+\frac{3s_{1}}{2}e^{\delta_{1}\rho/R}\dots+\frac{1}{4}\left(-3+\sqrt{5} \right)s_{2}\,e^{\delta_{2}\rho/R}+\dots\,, \notag\\
B&=\frac{1}{4}\ln\left(\frac{4}{3}\right)+s_{1}\,e^{\delta_{1}\rho/R}+\dots+s_{2}\,e^{\delta_{2}\rho/R}+\dots\,,\notag\\
U&=\frac{1}{4}\ln\left(\frac{4}{3}\right) -s_{1}\,e^{\delta_{1}\rho/R}+\dots +\left(2-\sqrt{5}\right)s_{2}\,e^{\delta_{2}\rho/R}+\dots\,,\notag\\
V&=-\frac{1}{4}\ln\left(\frac{4}{3}\right)-s_{1}\,e^{\delta_{1}\rho/R}+\dots +\left(-9+4\sqrt{5}\right)s_{2}\,e^{\delta_{2}
\rho/R}+\dots\,,
\end{align}
where $R=\frac{\sqrt{2}}{3^{3/4}}$, $\delta_{1}=2$ and $\delta_{2}=-1+\sqrt{5}$. This expansion corresponds to shooting out with two irrelevant 
operators of the $d=2$ IR SCFT of dimension $\Delta_{1}=4$ and $\Delta_{2}=1+\sqrt{5}$.

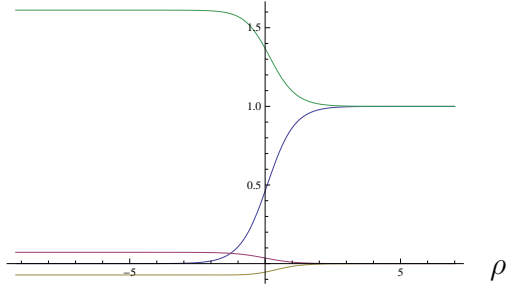
\begin{figure}[t!]
\centering
\begin{picture}(0.1,0.25)(0,0)
\put(190,5){\makebox(0,0){$\rho$}}
\end{picture}
\includegraphics[width=0.4\textwidth]{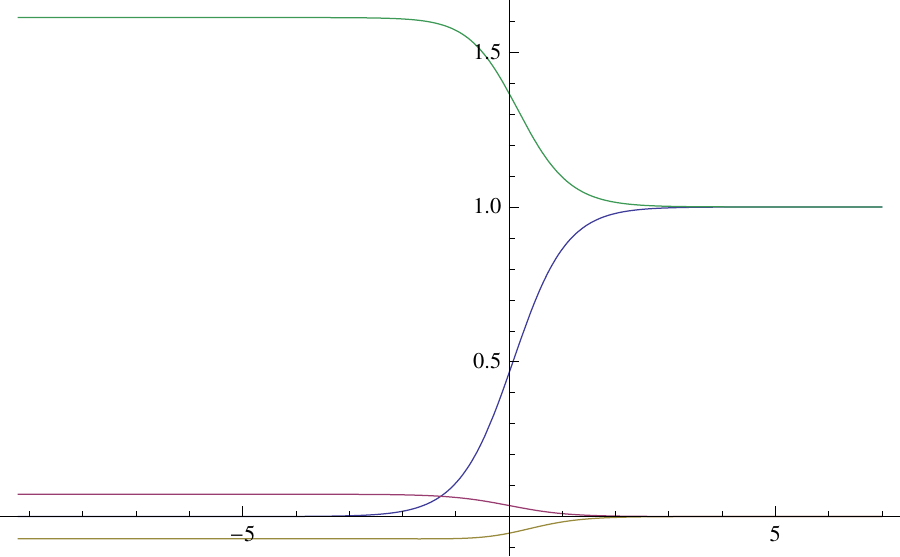}
\caption{Supersymmetric domain wall solutions interpolating between $AdS_5\times T^{1,1}$ and $AdS_3\times \mathbb{R}^2\times S^2\times S^3$ given
by \eqref{ansatz}.
From top to bottom in the figure, from the left, 
we have plotted the functions $A'$ (green), $U$ (red),  $B'$ (blue),
and $V$ (yellow).
\label{fig:dwall1}}
\end{figure}

Thus, we will obtain supersymmetric domain wall solutions interpolating between a deformation of $AdS_5\times T^{1,1}$ in the UV and the
$AdS_3\times \mathbb{R}^2\times S^2\times S^3$ solution \eqref{irfixed} in the IR, provided we can solve the differential equations
\eqref{eq:first_order_equations} (with $\lambda=2$), subject to the boundary conditions \eqref{uvexp}, \eqref{irexp1}.
Using a numerical shooting method we were able to match the two expansions provided that the constants take the values
\begin{align}
c_1=0.105\ldots\,,\quad
a_{0}=-0.130\ldots,\quad
s_{1}=-0.210\ldots\quad
s_{2}=0.480\ldots\,.
\end{align}
 In figure \ref{fig:dwall1} we have plotted the behaviour of the functions appearing in the supersymmetric domain wall solution.

\subsection{$D=5$ perspective and dual SCFT interpretation}\label{scftint}
The $AdS_5\times T^{1,1}$ solution is dual to an $N=1$ $d=4$ SCFT described in \cite{Klebanov:1998hh}. 
The global symmetry is $SU(2)\times SU(2)\times U(1)_R\times U(1)_B$ where $U(1)_R$ is the
$R$-symmetry and $U(1)_B$ is the baryonic symmetry. The field content includes
two gauge superfields, $W_1$ and $W_2$, corresponding to the $SU(N)\times SU(N)$ gauge group, as well
as bi-fundamental chiral fields. 
The UV expansion given in \eqref{uvexp} corresponds to deformations and expectation values of various operators
in the dual SCFT, whose precise details require a careful treatment of holographic renormalisation. However, it is not difficult to
extract the main features. 

We first 
observe that the domain wall flow solutions we have constructed are actually contained within
an $D=5$ $N=4$ gauged supergravity theory arising from a consistent truncation of the Kaluza-Klein reduction on
$T^{1,1}$. Recall that there is a consistent KK truncation of type IIB on a generic five-dimensional Sasaki-Einstein space to
a $D=5$ $N=4$ gauged supergravity with two $N=4$ vector multiplets \cite{Cassani:2010uw,Liu:2010sa,Gauntlett:2010vu,Skenderis:2010vz,Bah:2010cu,Liu:2010pq}. Expanding about the $AdS_5\times T^{1,1}$ vacuum
these fields give rise to $SU(2,2|1)$ multiplets, consisting of the gravity multiplet, a hypermultiplet, a massive gravition multiplet and a massive
vector multiplet. For the special case when $SE_5=T^{1,1}$, the $D=5$ $N=4$ gauged supergravity has an extra $N=4$ ``Betti"
vector multiplet \cite{Cassani:2010na,Bena:2010pr} (supersymmetry was discussed in \cite{Liu:2011dw}). Expanding about the $AdS_5\times T^{1,1}$ vacuum, the latter gives rise to a massless Betti vector multiplet, corresponding to the baryonic symmetry,
as well as a Betti hypermultiplet. In fact the solutions that we have just constructed 
are actually solutions of a further truncation to an $D=5$ $N=2$ gauged supergravity theory in which one discards the fields associated with the massive gravitino multiplet and also the massless Betti vector multiplet (while keeping the Betti hypermultiplet) \cite{Cassani:2010na}.

A key feature of the UV expansion
\eqref{uvexp} that is driving the supersymmetric flow , is that two $D=5$ ``axion" scalar fields (labelled by 
$b^\Phi$, $c^\Phi$ in \cite{Cassani:2010na} and $e_0^1,e^2_0$ in \cite{Liu:2011dw}) are equal to $-\lambda x^1, \lambda x^2$, 
respectively. 
These axions lie in the Betti hypermultiplet, which is identified with $Tr(W_1^2-W_2^2)$ in the dual SCFT \cite{Ceresole:1999zs,Cassani:2010na},
and are dual to marginal operators with dimension $\Delta=4$. 
The expansion \eqref{uvexp} also has a free
integration constant $c_1$ which corresponds to an operator of dimension $\Delta=6$ acquiring an expectation value. This scalar operator 
lies in the massive vector multiplet, which is identified with $Tr(W_1^2\bar W^2_1+W_2^2\bar W^2_2)+\dots$ in the dual SCFT 
\cite{Ceresole:1999zs,Cassani:2010na}.

It is worth highlighting the novelty of using $D=5$ axions in our construction.
Indeed the standard way of obtaining an
$AdS_3\times \mathbb{R}^2$ solution utilises massless $D=5$ vector fields carrying magnetic charges,
with field strength proportional to $vol(\mathbb{R}^2)$ \cite{Almuhairi:2011ws,Donos:2011pn}. 
By contrast in the solutions that we have constructed both
the R-symmetry vector field and the Betti vector field actually vanish identically\footnote{In section \ref{genflow} we will
construct more general solutions, still lying within the consistent truncation to $N=4$ $D=5$ gauged supergravity of \cite{Cassani:2010na,Bena:2010pr} but not the further truncation to $N=2$ of \cite{Cassani:2010na}, which have similar structure for the
axion fields and, amongst other features, also have a magnetic field for the Betti vector field.}.
The way in which the $D=5$ supersymmetry is being preserved for the solutions can be easily obtained using the results of
\cite{Liu:2011dw}\footnote{See equation (28)-(31) and especially (33) of \cite{Liu:2011dw}. The connection terms in (33) are potentially problematic, but our solutions have $A=\tau=\mathbb{A}=0$ and we note that $d\mathbb{A}$ is related to the linear axions via (15) of \cite{Liu:2011dw}.}

Finally, we note that in \cite{Singh:1998vf} a supersymmetric $AdS_3\times \mathbb{R}$
solution of $N=4$ $D=4$ gauged supergravity was constructed in which the $D=4$ axion field is linear in the coordinate on $\mathbb{R}$. This solution
was further discussed in \cite{Cowdall:1998bu}. It would be interesting to investigate the dimensional reduction of the $D=5$ gauged supergravity of
\cite{Cassani:2010na,Bena:2010pr} on a circle to $D=4$ and the relationship between the two $AdS_3$ solutions.

\subsection{Flux quantisation and central charge}\label{fluxq1}
In this section we analyse the flux quantisation for the supersymmetric domain wall solutions that we just constructed, assuming that 
we have compactified the two spatial directions labelled by $x_i$. The flux quantisation involves Page charges and is somewhat subtle; some additional details are presented in
appendix \ref{pchge}. We will also obtain the central charge of the
$d=2$ SCFT dual to the IR $AdS_3$ solution.

Note that so far we have been working in the Einstein frame with $\phi=0$. In this subsection, we view the metric as
being in the string frame and, furthermore, we redefine our R-R fields $F\to {g_s} F$ so that we are using similar conventions to \cite{Gukov:2004ym} 
although we will not set $2\pi l_s=1$ as they do.

We begin by assuming that the $x_i$ have period $2\pi d_i$,
\begin{align}
x_i=x_i+2\pi d_i\,,
\end{align}
and parametrise a $T^2$. The topology of internal space
is then $T^2\times S^2\times S^3$. The $S^2\times S^3$ is realised as a circle fibration over an $S^2_1\times S^2_2$ base
space. A positive orientation on $S^2\times S^3$ is given by $D\psi\wedge \mathrm{vol}_1\wedge \mathrm{vol}_2$.
A smooth three-manifold, $S^3_1$, that can be used to generate $H_3(S^2\times S^3,\mathbb{Z})$
is provided
by the circle bundle restricted to the $S^2_1$ factor on the base space.  We can also choose $S^3_2$, defined to be the 
circle bundle restricted to the $S^2_2$ factor on the base space, with opposite orientation. To find a smooth manifold that can be used to generate $H_2(S^2\times S^3,\mathbb{Z})$ we consider
any smooth manifold $S$ on the base that represents the cycle $[S]=[S^2_2]-[S^2_1]$. Since the circle bundle is trivial over $S$, there is a section $s$ and we can uses $s(S)$ to generate $H_2(S^2\times S^3,\mathbb{Z})$. 
A more detailed discussion is presented in appendix \ref{pchge}; here we record the values of the following integrals:
\begin{align}
&\int_{[S^3]} D\psi\wedge (\mathrm{vol}_1-\mathrm{vol}_2)=\int_{S^3_1} D\psi\wedge (\mathrm{vol}_1)=\int_{S^3_2} D\psi\wedge (-\mathrm{vol}_2)=16\pi^2\,,\nn
&-\int_{s(S)} \mathrm{vol}_1=\int_{s(S)} \mathrm{vol}_2=4\pi\,.
\end{align}

\subsubsection{Flux quantisation}
We begin with the three-form flux quantisation. We have
\begin{align}\label{handdc}
H=dB&=-\frac{L^2}{3}dx_1\wedge (\mathrm{vol}_1-\mathrm{vol}_2)\,,\nn
dC&=\frac{L^2}{3g_s}dx_2\wedge (\mathrm{vol}_1-\mathrm{vol}_2)\,,
\end{align}
and we note that both of these are globally defined and closed three-forms. We then
demand that
\begin{align}\label{qp}
\frac{1}{(2\pi l_s)^2}\int_{S^1_1\times s(S)} H&=\frac{4d_1}{3}\left(\frac{L}{l_s}\right)^2=Q_{N5}\in \mathbb{Z}\,,\nn
\frac{1}{(2\pi l_s)^2}\int_{S^1_2\times s(S)}dC^{(2)}&=-\frac{4d_2}{3g_s}\left(\frac{L}{l_s}\right)^2=Q_{D5}\in \mathbb{Z}\,.
\end{align}

Now we turn to the five-form. The relevant terms
are
\begin{align}
F&=\frac{L^4}{27 g_s}\left[
D\psi \wedge \mathrm{vol}_1 \wedge \mathrm{vol}_2+3 D\psi\wedge dx_1\wedge dx_2\wedge (\mathrm{vol}_1+\mathrm{vol}_2) \right]+\dots\,,
\end{align}
which is globally defined since the one-form $D\psi$ is.
Recall that the Bianchi identity for the five-form is given by $dF-H\wedge dC=0$. We will demand that
a corresponding Page charge should be quantised. Specifically we demand that
\begin{align}\label{ffq}
\frac{1}{(2\pi l_s)^4}\int_{\Sigma_5}( F-B\wedge dC)
\in \mathbb{Z}\,,
\end{align}
for any five-cycle $\Sigma_5$. As we will see there are some subtleties in imposing this condition.
Furthermore, as will be clear from the subsequent discussion, the subtleties are not removed by having the Page charges defined by integrating, instead,
the five-form $F+C\wedge dB$, for example.

There are two five-cycles to consider. For $\Sigma_5=S^2\times S^3$ the gauge-dependent terms involving the two-form $B$
do not contribute and
we find
\begin{align}\label{Nval}
N\equiv \left(\frac{L}{l_s}\right)^4\frac{vol(T^{1,1})}{g_s4\pi^4}\in \mathbb{Z}\,,
\end{align}
where\footnote{Recall that the central charge of the
$d=4$ SCFT dual to $AdS_5\times T^{1,1}$ is given by $a=(N^2/4)\pi^3/( vol(T^{1,1})$.} $vol(T^{1,1})= 16\pi^3/27$. 

The delicate case to consider is the five-cycle that is the product of the $T^2$ with the generator of $H_3(S^2\times S^3)$.
Recall that for the latter we can consider $S^3_1$ which is the circle bundle over $S^2_1$ at any fixed point on $S^2_2$.
We can also consider $S^3_2$ which is the circle bundle over $S^2_2$ at any fixed point on $S^2_1$, but with opposite orientation. 
We first calculate that
\begin{align}
&\frac{1}{(2\pi l_s)^4}\int_{T^2\times S^3_1} F=\left(\frac{L}{l_s}\right)^4\frac{4d_1 d_2}{9g_s}\,,\nn
&\frac{1}{(2\pi l_s)^4}\int_{T^2\times S^3_2} F=-\left(\frac{L}{l_s}\right)^4\frac{4d_1 d_2}{9g_s}\,.
\end{align}
These differ because $F$ is not closed and hence does not define a cohomology class.

We now need to consider a suitable gauge for the two-form $B$.
It does not seem possible to find a single gauge-choice for $B$ that is well defined as a two-form for an arbitrary
three manifold representing $H_3(S^2\times S^3)$. However, it is possible to find a gauge for a specific representative.
In particular, if we integrate over $S^3_1$ we can choose the gauge
$B^{(1)}=\frac{L^2}{3}dx_1\wedge (d\psi+P_1-P_2)$, where $dP_i=\mathrm{vol}_i$,
while 
if we integrate over $S^3_2$ we can choose a different gauge
$B^{(2)}=\frac{L^2}{3}dx_1\wedge (-d\psi+P_1-P_2)$. We will discuss this more carefully below.
We then calculate
\begin{align}\label{calcb}
&\frac{1}{(2\pi l_s)^4}\int_{T^2\times S^3_1} -B^{(1)}\wedge dC=\left(\frac{L}{l_s}\right)^4\frac{4d_1 d_2}{9g_s}\,,\nn
&\frac{1}{(2\pi l_s)^4}\int_{T^2\times S^3_2} -B^{(2)}\wedge dC=-\left(\frac{L}{l_s}\right)^4\frac{4d_1 d_2}{9g_s}\,.
\end{align}

The quantisation condition that we will impose is given by 
\begin{align}\label{qc}
\bar N\equiv \left(\frac{L}{l_s}\right)^4\frac{8d_1 d_2}{9g_s}\in \mathbb{Z}\,.
\end{align}
With this condition we see that the Page charge \eqref{ffq} when $\Sigma_5=T^2\times S^3_1$ and 
$\Sigma_5=T^2\times S^3_2$, with the gauge-choices for $B$ given above, are equal to $\bar N$ and -$\bar N$, respectively.
While these are quantised, one might be concerned that they are not equal given the two choices of $\Sigma_5$ are homologous and
that the integrand is closed. The key point is that the integrand is not a differential form since it changes under gauge-transformations, 
which we make precise below. 

Note that the condition \eqref{qc} is equivalent to the statement that the product of the three-form fluxes is constrained to be an even number:
\begin{align}\label{conchges}
2\bar N= -Q_{N_5} Q_{D_5}\,.
\end{align}

Let us now elaborate a little on the gauge choices for $B$ that we made above. We first introduce four coordinate patches
$U_{NN},U_{NS},U_{SN},U_{SS}$, each isomorphic to $\mathbb{R}^4\times S^1$,
to cover $S^2\times S^3$.
We take $U_{NN}$ 
to consist of the northern hemispheres of the two $S^2$'s on the base as well as a coordinate $\psi_{NN}$ with period $4\pi$.
Next, $U_{NS}$ is the northern hemisphere of $S^2_1$ and the southern hemisphere of $S^2_2$ on the base, 
as well as a coordinate $\psi_{NS}$ with period $4\pi$, and similarly for the rest. Now we know that the one-form 
$D\psi\equiv d\psi+P$ is globally defined and we have
\begin{align}
D\psi
&=d\psi_{NN}+(1-\cos\theta_1)d\phi_1+(1-\cos\theta_2)d\phi_2\,,\nn
&=d\psi_{NS}+(1-\cos\theta_1)d\phi_1+(-1-\cos\theta_2)d\phi_2\,,\nn
&=d\psi_{SN}+(-1-\cos\theta_1)d\phi_1+(1-\cos\theta_2)d\phi_2\,,\nn
&=d\psi_{SS}+(-1-\cos\theta_1)d\phi_1+(-1-\cos\theta_2)d\phi_2\,.
\end{align}
On the overlaps of the patches we have
\begin{align}\label{patchcb}
\psi_{NN}=\psi_{NS}-2\phi_2=\psi_{SN}-2\phi_1=\psi_{SS}-2\phi_1-2\phi_2\,,
\end{align}
which shows that we have a good circle bundle: e.g. $\psi_{NN}/2=\psi_{NS}/2+ie^{-i\phi_2}d(e^{i\phi_2})$ (and we note that the factors of $1/2$ are present because $\psi$ has period $4\pi$).

For the five manifold $T^2\times S^2\times S^3$ we can consider four coordinate patches, isomorphic to $T^2\times \mathbb{R}^4\times S^1$, labelled
in the same way. In particular, as we will see, the $T^2$ essentially just comes along for the ride.
Now we consider the gauge for the two-form $B$ given in the $NN$ patch by
\begin{align}\label{gaone}
B^{(1)}=\frac{L^2}{3}dx_1\wedge (d\psi_{NN}+(1-\cos\theta_1)d\phi_1-(1-\cos\theta_2)d\phi_2)\,,
\end{align}
which is clearly well defined in $U_{NN}$. We see that it is also well defined in $U_{SN}$, after using \eqref{patchcb}. Thus, it makes sense
to integrate this over $S^3_1$ which lies in the union of these two patches\footnote{Observe that we have defined $S^3_1$ here 
to be sitting at a fixed point on the northern hemisphere of the second two-sphere.}, giving the result in \eqref{calcb}.
Observe that if we instead move to $U_{NS}$ then we have 
\begin{align}
B^{(1)}=\frac{L^2}{3}dx_1\wedge (d\psi_{NS}+(1-\cos\theta_1)d\phi_1+(1+\cos\theta_2)d\phi_2) -\frac{4L^2}{3}dx_1\wedge d\phi_2\,.
\end{align}
Now the first term on the right hand side is well defined in this patch, but the last term isn't. 
However, moving to this patch we can employ
a gauge-transformation on the two-form given by
\begin{align}
\delta B&=\frac{4L^2}{3}dx_1\wedge d\phi_2\,.
\end{align}
To see this is well defined, we recall that the definition of the integrality of the three-form curvature $\bar H$ of a gerbe\footnote{As we will see in appendix \ref{pchge} the essential aspects of these arguments don't really involve gerbes but more familiar gauge-connections.} 
connection (or ``curving")  $\bar B$
is given by $\frac{1}{2\pi}\int \bar H\in\mathbb{Z}$, and so we should absorb a factor of $2\pi l_s^2$ in $B$ and $H$ and consider $\frac{1}{2\pi l_s^2}\delta B$.
Using the flux quantisation condition \eqref{qp} we find
\begin{align}
 \frac{1}{2\pi l_s^2}\delta B&=\frac{Q_{N_5}}{2\pi d_1}{dx_1}\wedge d\phi_2\,,\nn
&=-\frac{dx_1}{2\pi d_1}\wedge (ie^{-iQ_{N_5}\phi_2}de^{iQ_{N_5}\phi_2})\,,
\end{align}
which is indeed a bona-fide gauge-transformation for the gerbe. 
Thus we have shown that $B^{(1)}$ patches together to properly define a conenction for the gerbe with curvature $H$,
and furthermore $B^{(1)}$ gives a well defined two-form on $S^{3}_1$ and hence can be integrated over it.

Similarly, if we consider the gauge for the two-form $B$ given by
\begin{align}\label{gatwo}
B^{(2)}=\frac{L^2}{3g_s}dx_1\wedge (-d\psi_{NN}+(1-\cos\theta_1)d\phi_1-(1-\cos\theta_2)d\phi_2)\,.
\end{align}
we see that it is well defined in $U_{NN}$ and also on $U_{NS}$. Hence this is something that can be integrated on the manifold $S^3_2$ (sitting at a point in the northern hemisphere of the first two sphere) leading
to the result given in \eqref{calcb}. To see that this is a well-defined gerbe connection we can calculate the difference
between $B^{(1)}$ and $B^{(2)}$ on, say, the
$NN$ patch. We find
\begin{align}
 \frac{1}{2\pi l_s^2}(B^{(1)}-B^{(2)})&=\frac{dx_1}{2\pi d_1}\wedge (ie^{iQ_{N_5}\psi_{NN}/2}de^{-iQ_{N_5}\psi_{NN}/2})\,.
 \end{align}
which is a good gauge-transformation since $\psi_{NN}$ has period $4\pi$.

\subsubsection{Central charge}\label{ccc}
These flux quantisation conditions we have just derived are valid for the entire domain wall flow solution. We can also calculate the
central charge of the $d=2$ $(0,2)$ SCFT that is dual to the $AdS_3$ solution \eqref{ads3fp}. We use the standard formula
\begin{align}\label{cchgeb}
c=\frac{3R_{AdS_3}}{2G_3}\,,
\end{align}
where $R_{AdS_3}$ is the $AdS_3$ radius and $G_3$ is the effective 3d Newton's constant.
Our D=10 Lagrangian in the string frame is of the form
\begin{align}\label{actconv}
\frac{1}{(2\pi)^7g_s^2 l_s^8}\sqrt{-g}e^{-2\phi}R+\dots\,,
\end{align}
and a calculation leads to
\begin{align}\label{ccfg}
c&=\frac{3}{2}| N Q_{N5} Q_{D5}|\,,\nn
&= 3| N\bar N|\,.
\end{align}
where the second expression arises from \eqref{conchges}.

\subsection{T-duality and enhanced supersymmetry}\label{enhsus}
It was pointed out in \cite{Donos:2008ug} that, locally, the $AdS_3\times T^2\times S^2\times S^3$ IR
solution \eqref{irfixed} is related, after
two T-dualities on the $T^2$, to the well known $AdS_3\times S^3\times S^3\times S^1$ solution of type IIB that is dual
to a $d=2$ SCFT with large $(4,4)$ supersymmetry \cite{Cowdall:1998bu}
(see \cite{Gukov:2004ym} for a detailed discussion). 
To see this we carry out two T-dualities along the directions $z_1,z_2$ (as in \eqref{irfixed}) using, for example, 
the formulae in appendix B of \cite{Donos:2008ug}. After then
introducing rescaled coordinates 
\begin{align}\label{pers}
\bar z_1=\frac{2^{1/2}3^{3/2}}{L^2} z_1,\qquad 
\bar z_2=\frac{3^{3/2}}{2^{1/2}L^2}z_2\,,
\end{align}
and defining
\begin{align}\label{coordchange}
\alpha_1=\frac{1}{2}(\psi-\bar z_1),\qquad 
\alpha_2=\frac{1}{2}(\psi+\bar z_1)\,,
\end{align}
we obtain
\begin{align}
ds^2&=\frac{2L^2}{3^{3/2}}\left[ds^2(AdS_3)+2 ds^2(S^3_1)+2 ds^2(S^3_2)+d\bar z_2^2\right]\,,\nn
dC^2&=e^{-\phi_0}\frac{2L^2}{3^{3/2}}\left[2 \mathrm{vol}(AdS_3) +4 \mathrm{vol}(S^3_1)+4 \mathrm{vol}(S^3_2)\right]\,,\nn
e^{\phi_0}&=\frac{3^{3/2}}{L^2}\,,
\end{align}
where 
\begin{align}\label{gog}
ds^2(S^3_i)=\frac{1}{4}\left[ds^2_i+(d\alpha_i+P_i)^2\right]\,.
\end{align}
If $\alpha_i$ are periodic coordinates with period $4\pi$ then \eqref{gog} is the metric on a round, unit radius 
three-sphere
and we have the standard $AdS_3\times S^3\times S^3\times S^1$ solution of type IIB supergravity, which we 
we briefly review in appendix \ref{app:A}. 

Observe that \eqref{coordchange} implies that 
\begin{align}
\partial_\psi=\frac{1}{2}(\partial_{\alpha_1}+\partial_{\alpha_2}),\qquad
\partial_{\bar z_1}=\frac{1}{2}(-\partial_{\alpha_1}+\partial_{\alpha_2})\,.
\end{align}
Thus, locally, starting with the 
$AdS_3\times S^3\times S^3\times S^1$ solution we can obtain the $AdS_3\times T^2\times S^2\times S^3$
solution by carrying out a T-duality on the $S^1$ factor, generated by $\partial_{\bar z}$, 
and on the diagonal of the two $U(1)$ Hopf fibres
generated by $\frac{1}{2}(-\partial_{\alpha_1}+\partial_{\alpha_2})$.
Recalling that the $AdS_3\times S^3\times S^3\times S^1$ solution preserves 16 supersymmetries (8 Poincar\'e and 8 superconformal), this
suggests that the $AdS_3\times T^2\times S^2\times S^3$ solution will preserve more than the obvious 4 supersymmetries.
Indeed the explicit Killing spinors for the $AdS_3\times S^3\times S^3\times S^1$ solution, in a $D=11$ incarnation, were constructed
in \cite{Gauntlett:1998kc} and the corresponding superisometry algebra was found. Using the arguments
in section 7 of \cite{Gauntlett:1998kc} one can determine the Killing spinors which are
left invariant under the action of the Lie derivative with respect to the Killing vector generating the diagonal $U(1)$ on the  $S^3\times S^3$ factor.
We find that this action preserves all eight Killing spinors given by equation (48) of \cite{Gauntlett:1998kc} and four of the eight given by equation (47) of \cite{Gauntlett:1998kc} (in particular
satisfying the projection $(1+\Gamma^{121'2'})\varepsilon=0$ in the notation of that paper).
We have verified this counting by a direct construction of the Killing spinors for the type IIB solutions
in appendix \ref{appes}.

Using the results of \cite{Gauntlett:1998kc} we can also deduce the superisometry algebra. It will
be of the form $ D(2,1|\alpha)\times G$, with $G\subset D(2,1|\alpha)$ and the two factors having
bosonic sub-algebras given by $SL(2)\times SU(2)\times SU(2)$ and 
$SL(2)\times U(1)^2$, respectively.

A more careful examination of the global aspects of the T-duality will be left to future work. Note that
the relevant $AdS_3\times S^3\times S^3\times S^1$ solution is fixed by two integers, $Q_{D1}, Q_{D5}$ and the central charge is given by $c=3 Q_{D1} Q_{D5}$, which can be compared with the second expression in
\eqref{ccfg}. However, the first expression in \eqref{ccfg} suggests that orbifolds of $AdS_3\times S^3\times S^3\times S^1$ might need to be considered.

\section{A more general class of flows}\label{genflow}
In this section we construct a more general class of flows interpolating between $AdS_5\times T^{1,1}$ 
and a one-parameter family of $AdS_3\times \mathbb{R}^2\times S^2\times S^3$ solutions found in \cite{Donos:2008ug}.
The flows again preserve $(0,2)$ supersymmetry and we show in appendix \ref{appes} that the $AdS_3$ fixed
point solutions preserve $(4,2)$ superconformal symmetry.

Specifically, here we
consider an ansatz for the type IIB fields given by
\begin{align}\label{eq:gen_ansatz}
\frac{1}{L^2}ds^{2}&= e^{2A}\,\left(-dt^{2}+dx^{2}\right)+e^{2B}\,\left(dx_{1}^{2}+dx_{2}^{2}\right)+d\rho^{2}+
\frac{1}{6}\left(e^{2U_{1}}\,ds_{1}^{2}+e^{2U_{2}}\,ds_{2}^{2}\right) +e^{2V}\,\eta^{2}\,,\notag\\
\frac{1}{L^4}F&=4e^{2A+2B-V-2U_{1}-2U_{2}}\,dt\wedge dx\wedge dx_{1}\wedge dx_{2}\wedge d\rho+\frac{1}{9}\,\eta\wedge \mathrm{vol}_{1}\wedge \mathrm{vol}_{2}\notag\\
&+dx_{1}\wedge dx_{2}\wedge \eta\wedge \left[\frac{\left(\lambda^{2}-4f^2 \right)}{12}\,\left(\mathrm{vol}_{1}+\mathrm{vol}_{2}\right)-\frac{\lambda \left(f+Q\lambda\right)}{3}\,\left(\mathrm{vol}_{1}-\mathrm{vol}_{2} \right)\right]\notag\\
&+e^{2A-2B-V}\frac{\left(\lambda^2-4f^{2}\right)}{12}\,dt\wedge dx\wedge d\rho\wedge \left(e^{2U_{2}-2U_{1}}\mathrm{vol}_{2}+e^{2U_{1}-2U_{2}}\mathrm{vol}_{1}\right)\notag\\
&-e^{2A-2B-V}\frac{\lambda \left(f+Q\lambda\right)}{3}\,dt\wedge dx\wedge d\rho\wedge\left(e^{2U_{2}-2U_{1}}\mathrm{vol}_{2}-e^{2U_{1}-2U_{2}}\mathrm{vol}_{1} \right)\,, \notag\\
\frac{1}{L^2}G&=(dx_1-idx_2)\wedge\left(\frac{\lambda}{6}\left(\mathrm{vol}_{1}-\mathrm{vol}_{2} \right)+d\left(f\,\eta\right)\right)\,,\notag\\
&=(dx_1-idx_2)\wedge\left(\frac{\lambda}{6}\left(\mathrm{vol}_{1}-\mathrm{vol}_{2} \right)+ f^{\prime} d\rho\wedge\eta+\frac{f}{3}\left(\mathrm{vol}_{1}+\mathrm{vol}_{2}\right)\right)\,,
\end{align}
with $A,B,U_1,U_2,V,f$ all functions of $\rho$, and $\lambda,Q$ are constants. The interpretation within the dual $D=4$ SCFT will be discussed below.
One can check that the five-form is self-dual and that the Bianchi identities \eqref{bianch} are satisfied.
 
We find that if we again write the Killing spinors as $\varepsilon=e^{A/2}\varepsilon_0$, demand that they satisfy the projections 
\eqref{eq:projections}, \eqref{proj2} and \eqref{second}, then the Killing spinor equations \eqref{eq:kill1},\eqref{eq:kill2}
lead to the following system of first order differential equations
\begin{align}\label{eq:eom_gen}
&A^{\prime}-e^{-V-2U_{1}-2U_{2}}+\frac{4f^{2}-\lambda^{2}}{8}e^{-2B-V}\,\left(e^{-2U_{1}}+e^{-2U_{2}}\right)
\notag\\&
\qquad\qquad\qquad\qquad\qquad+\frac{\lambda \left(f+Q\lambda\right)}{2}e^{-2B-V}\,\left(e^{-2U_{1}}-e^{-2U_{2}} \right)=0\,,\notag\\
&B^{\prime}-e^{-V-2U_{1}-2U_{2}}-\frac{4f^{2}-\lambda^{2}}{8}e^{-2B-V}\,\left(e^{-2U_{1}}+e^{-2U_{2}}\right)\notag\\
&\qquad\qquad\qquad\qquad\qquad-\frac{\lambda \left(f+Q\lambda\right)}{2}e^{-2B-V}\,\left(e^{-2U_{1}}-e^{-2U_{2}} \right)=0\,,\notag\\
&U_{1}^{\prime}-e^{V-2U_{1}}+e^{-V-2U_{1}-2U_{2}}-\frac{4f^{2}-\lambda^{2}}{8}\,e^{-2B-V}\,\left(e^{-2U_{1}} -e^{-2U_{2}}\right)\notag\\
&\qquad\qquad\qquad\qquad\qquad-\frac{\lambda \left(f+Q\lambda\right)}{2}e^{-2B-V}\,\left(e^{-2U_{1}}+e^{-2U_{2}}\right)=0\,,\notag\\
&U_{2}^{\prime}-e^{V-2U_{2}}+e^{-V-2U_{1}-2U_{2}}+\frac{4f^{2}-\lambda^{2}}{8}\,e^{-2B-V}\,\left(e^{-2U_{1}} -e^{-2U_{2}}\right)\notag\\
&\qquad\qquad\qquad\qquad\qquad+\frac{\lambda \left(f+Q\lambda\right)}{2}e^{-2B-V}\,\left(e^{-2U_{1}}+e^{-2U_{2}}\right)=0\,,\nn
&V^{\prime}-3e^{-V}+e^{V-2U_{1}}+e^{V-2U_{2}} +e^{-V-2U_{1}-2U_{2}}-\frac{4f^{2}-\lambda^{2}}{8}e^{-2B-V}\,\left(e^{-2U_{1}}+e^{-2U_{2}} \right)\notag\\
&\qquad\qquad\qquad\qquad\qquad-\frac{\lambda \left(f+Q\lambda\right)}{2}e^{-2B-V}\,\left(e^{-2U_{1}}-e^{-2U_{2}}\right)=0\,,\notag
\end{align}
\begin{align}
&f^{\prime}+2f\,\left(e^{V-2U_{1}}+e^{V-2U_{2}}\right)+\lambda\,\left(e^{V-2U_{1}}-e^{V-2U_{2}}\right)=0\,.
\end{align}
The first five equations come from \eqref{eq:kill1} while the sixth comes from \eqref{eq:kill2}.
One can show that these equations imply that the equation of motion for the complex three-form $G$, given in \eqref{Geom}, is satisfied.

Notice that if we set $U_1=U_2=U$ and $f=Q=0$ we recover the equations \eqref{eq:first_order_equations} that we had in the last section\footnote{Observe that if we set $f=0$, with $\lambda\ne 0$, we must have $U_1=U_2=U$ and $Q=0$.}.
In particular, the $AdS_5\times T^{1,1}$ solution is obtained via \eqref{ads5fp}. The set of equations \eqref{eq:eom_gen} also
admits the following one parameter family of $AdS_3$ solutions
\begin{align}\label{eq:AdS3_family}
A&=\rho/R\equiv \frac{3^{3/4}}{\sqrt{2}\,\left(1-4 Q^{2}\right)^{1/4}}\rho,\qquad B=
b_{0}\equiv\frac{1}{4}\,\ln\left[\frac{\lambda^4}{12}\,\left(1-4Q^{2}\right) \right]\,,\notag\\
U_{1}&=u_{1}\equiv\frac{1}{4}\,\ln\left[\frac{4}{3}\frac{1-2Q}{1+2Q} \right],\quad\qquad U_{2}=u_{2}\equiv\frac{1}{4}\,\ln\left[\frac{4}{3}\frac{1+2Q}{1-2Q} \right]\,,\nn
V&=v\equiv\frac{1}{4}\,\ln\left[\frac{3}{4}\,\left(1-4Q^{2}\right) \right],\qquad f=-\lambda Q\,.
\end{align}
with 
\begin{align}
0\leq Q<1/2\,.
\end{align}
After scaling $x_{i}=\frac{2^{1/2}}{\lambda 3^{1/2}(1-4Q^2)^{1/4}}z_{i}$ the resulting type IIB solution can be written as
\begin{align}\label{fpfg}
\frac{1}{L^2}ds^{2}&=\frac{\left(1-4Q^{2}\right)^{\frac{1}{2}}}{3^{3/2}}\,\left(2\,ds^{2}\left(AdS_{3}\right)
+\frac{1}{\left(1-4Q^{2}\right)^{\frac{1}{2}}}\,\left(dz_{1}^{2}+dz_{2}^{2}\right)\right.\notag\\
&\left. \qquad\qquad\qquad\qquad\qquad+\frac{1}{1+2Q}d{s}^{2}_{1}+\frac{1}{1-2Q}d{s}_{2}^{2}+\frac{1}{2}\,\left(d\psi+P \right)^{2} \right)\,,\nn
\frac{1}{L^4}F&=\frac{1}{27}\Bigg\{  \mathrm{vol}(AdS_3)\left[4(1-4Q^2)^{1/2}dz_1\wedge dz_2+
2 (1-2Q)^2\mathrm{vol}_1+2(1+2Q)^2\mathrm{vol}_2\right]  \nn
&\qquad+\left(d\psi+P \right)\wedge\left[\mathrm{vol}_1 \wedge \mathrm{vol}_2+ \frac{(1-4Q^2)^{1/2}}{2} dz_1\wedge dz_2\wedge (\mathrm{vol}_1+\mathrm{vol}_2) \right]\Bigg\}\,,\nn
\frac{1}{L^2}G&=\frac{1}{3^{3/2}2^{1/2}(1-4Q^2)^{1/4}}(dz_1-idz_2)\wedge\left[ (1-2Q)\mathrm{vol}_1-(1+2Q)\mathrm{vol}_2\right]\,.
\end{align}
which is precisely the same one-parameter family of solutions\footnote{We should identify
$L^2=[3^{3/2}(l_1+l_2)^{1/2}/2(l_1l_2)^{1/2}]L^2_{there}$, $Q=(l_1-l_2)/2(l_1+l_2)$, $(z_1-i z_2)=(l_1l_2)^{1/4}u$ and
$\psi=z$.} found in section 3.1.2 of \cite{Donos:2008ug}. 

\subsection{The supersymmetric flows}
We now discuss the domain wall solutions that interpolate between $AdS_5\times T^{1,1}$ in the UV and this one-parameter
family of $AdS_3\times \mathbb{R}^2\times S^2\times S^3$ solutions. There is an expansion about
the $AdS_5\times T^{1,1}$ solution 
that involves three integration constants $c_i$, in addition to $Q$ and the deformation parameter $\lambda$.
The UV expansion analogous to \eqref{uvexp} is rather long so we shall not write it out explicitly. The key feature is that as $\rho\to\infty$
three integration constants $c_i$ appear, schematically, as
\begin{align}\label{uvexp2}
U_{1}&=c_{2}\,e^{-2 \rho}+\dots +c_{1}\,e^{-6 \rho}+\notag\cdots\\
U_{2}&=-c_{2}\,e^{-2 \rho}+\dots+c_{1}\,e^{-6 \rho}\notag+\cdots\\
V&=\dots-4c_{1}\,e^{-6 \rho}\notag+\cdots\\
f&=\dots+c_{3}e^{-4\rho}+\cdots
\end{align}
The UV expansion that we had before, given in \eqref{uvexp}, is obtained by setting $c_2=c_3=0$.
We will discuss the holographic interpretation of the $c_i$ below.

We next develop an expansion about the $AdS_3$ solution \eqref{eq:AdS3_family} in the IR. We find that as $\rho\to -\infty$ it can be constructed from
four constants $s_1,s_2,s_3$ and $a_0$:
\begin{align}
A&=a_{0}+\rho/R+\frac{3}{2}s_{1}\,e^{\delta_{1}\rho/R}+w_{1}\,s_{2}\,e^{\delta_{2}\rho/R}+w_{2}\,s_{3}\,e^{\delta_{3}\rho/R}+\cdots\,,\notag\\
B&=b_{0}+s_{1}\,e^{\delta_{1}\rho/R}+s_{2}\,e^{\delta_{2}\rho/R}+s_{3}\,e^{\delta_{3}\rho/R}+\cdots\,,\notag\\
U_{1}&=u_{1}-s_{1}\,e^{\delta_{1}\rho/R}-\,s_{2}\,e^{\delta_{2}\rho/R}+w_{3}\,s_{3}\,e^{\delta_{3}\rho/R}+\cdots\,,\notag\\
U_{2}&=u_{2}-s_{1}\,e^{\delta_{1}\rho/R}+w_{4}\,s_{2}\,e^{\delta_{2}\rho/R}-s_{3}\,e^{\delta_{3}\rho/R}+\cdots\,,\notag\\
f&=-\lambda Q+w_{5}\,s_{2}\,e^{\delta_{1}\rho/R}+w_{6}\,s_{3}e^{\delta_{3}\rho/R}+\cdots\,,\notag\\
V&=v-s_{1}\,e^{\delta_{1}\rho/R}+w_{7}\,s_{2}\,e^{\delta_{2}\rho/R}+w_{8}\,s_{3}\,e^{\delta_{3}\rho/R}+\cdots\,,
\end{align}
where $\delta_{1}=2$, $\delta_{2}=-1+\sqrt{5-8Q}$ and $\delta_{3}=-1+\sqrt{5+8Q}$ and $w_{i}$ are functions of $Q$.
Explicitly we have
\begin{align}
&w_1=\tfrac{-3+8Q+\sqrt{5-8Q}}{(4-8Q)},\quad
w_2=\tfrac{-3-8Q+\sqrt{5-8Q}}{(4+8Q)},\quad
w_3=\tfrac{-5-2Q+2\sqrt{5-8Q}}{(-1+2Q)},\quad w_4=\tfrac{5-2Q-2\sqrt{5-8Q}}{(1+2Q)},\nn
&w_5=-2\lambda(-2+2Q+\sqrt{5-8Q}),\quad
w_6=2\lambda(-2-2Q+\sqrt{5-8Q})\,,\nn
&w_7=\tfrac{-9+6Q+4\sqrt{5-8Q}}{(1+2Q)},\quad
w_8=\tfrac{9+6Q-4\sqrt{5-8Q}}{(-1+2Q)}.
\end{align}
This expansion corresponds to shooting out with three IR irrelevant 
operators of dimension $\Delta_{1}=4$,
$\Delta_{2}=1+\sqrt{5-8Q}$ and $\Delta_{3}=1+\sqrt{5+8Q}$. 
Observe that if we set $Q=0$ and in addition we also set $s_2=s_3$ then we recover the expansion \eqref{irexp1}
that we had in the last section. It is worth emphasising that when $Q=0$ the enlarged ansatz of this section, with $U_1\ne U_2$ and $f\ne 0$, 
leads to an extra irrelevant IR operator parametrised by, say, $s_2-s_3$.

We now set $\lambda=2$. Fixing $0\le Q<1/2$, our UV expansion has three integration constants and our IR expansion has four.
On the other hand our system of differential equations \eqref{eq:eom_gen} is fixed by six integration constants. Thus, for each value of $Q$,
including $Q=0$, we expect to have a one parameter family of supersymmetric flows connecting the deformed 
$AdS_5\times T^{1,1}$ solution with the corresponding $AdS_3\times \mathbb{R}^2\times S^2\times S^3$
solution. We have constructed a couple of examples of such flows numerically, including when $Q=0$. The $Q=0$ solutions of the last section are distinguished in this family by having $U_1=U_2$, or equivalently $c_2=0$ in the UV expansion \eqref{uvexp2}, which is associated with a particular relevant operator with $\Delta=2$ being switched off (see below).
In the next subsection we will argue that flux quantisation implies a rationality condition on $Q$. 

The more general ansatz \eqref{eq:gen_ansatz} that we are using for the supersymmetric flows is again contained
within the $N=4$ $D=5$ gauged supergravity obtained from a consistent Kaluza-Klein truncation on $T^{1,1}$
\cite{Cassani:2010na,Bena:2010pr,Liu:2011dw}.
As before, the two $D=5$ axion scalar fields in the Betti hypermultiplet,
dual to $\Delta=4$ operators, are equal to $-\lambda x^1, \lambda x^2$, respectively, and this deformation is 
driving the flow. We next note that the field strength of the massless vector field lying in the Betti vector multiplet 
(labelled $d(a_1^\Phi)$ in \cite{Cassani:2010na} and $r_2$ in \cite{Liu:2011dw}) is of 
the form $\lambda(f+Q\lambda)dx_1\wedge dx_2$. This reveals that the $Q$-deformation corresponds to switching on 
a magnetic field for the massless gauge-field dual to the $\Delta=3$ current associated with the baryonic $U(1)$ symmetry.
The three constants $c_i$ in \eqref{uvexp2} 
are related to various operators in the dual $d=4$ SCFT acquiring expectation values, which can be deduced from
the results of \cite{Gauntlett:2010vu,Cassani:2010na,Liu:2011dw}. 
The constant $c_1$ is again associated with the scalar in the massive vector multiplet which is dual to an operator
of dimension $\Delta=6$. Similarly, the constant  
$c_2$ is associated with the scalar in the Betti vector multiplet (labelled $w$ in \cite{Cassani:2010na}) dual to an operator
of dimension $\Delta=2$.  Finally, 
the constant $c_3$ is associated with the massive one-forms (labelled $b_1,c_1$ in \cite{Cassani:2010na})
appearing in the massive gravitino multiplet, dual to operators with dimension $\Delta=5$.

\subsection{Flux quantisation and central charge}

Repeating the steps in section \ref{fluxq1} (with $\lambda=2$) we find that 
the quantisation of the three-form flux leads to the same results, namely
\begin{align}\label{qpsb}
Q_{N5}=\frac{4d_1}{3}\left(\frac{L}{l_s}\right)^2\in \mathbb{Z}\,,\nn
Q_{D5}=-\frac{4d_2}{3g_s}\left(\frac{L}{l_s}\right)^2\in \mathbb{Z}\,.
\end{align}
Similarly, integrating the five-from flux on the five-cycle $\Sigma_5=S^2\times S^3$
implies
\begin{align}\label{Nvalgenap}
N\equiv \left(\frac{L}{l_s}\right)^4\frac{4}{27g_s\pi }\in \mathbb{Z}\,,
\end{align}
as before. 

The calculation of the Page charge for the five-cycle $\Sigma_5=T^2\times S^3$, however,
exhibits some new features. We follow the same prescription that we deployed in section 2.
To carry out the integral \eqref{ffq}
over $T^2\times S^3_1$ and $T^2\times S^3_2$ we use the gauge choices:
\begin{align}
B^{(1)}=\frac{L^2}{3}dx_1\wedge [(d\psi+P_1-P_2)+\frac{f}{3}D\psi],\nn
B^{(2)}=\frac{L^2}{3}dx_1\wedge [(-d\psi+P_1-P_2)+\frac{f}{3}D\psi],
\end{align}
respectively, and we obtain the two quantisation conditions
\begin{align}\label{noneandntwo}
\bar N_1\equiv \left(\frac{L}{l_s}\right)^4\frac{d_1 d_2}{9g_s}(8-4f^2-4f-16Q)\in \mathbb{Z}\,,\nn
\bar N_2\equiv -\left(\frac{L}{l_s}\right)^4\frac{d_1 d_2}{9g_s}(8-4f^2+4f+16Q)\in \mathbb{Z}\,,
\end{align}
respectively.

Since $f$ is a function of $\rho$ we obviously cannot satisfy \eqref{noneandntwo} throughout the whole flow. 
We can however, demand that the flux is properly quantised at the $AdS_5$ boundary, where $f\to 0$, 
and also at the $AdS_3$ fixed point, where $f=-2Q$, for suitable choices of rational $Q$. This would place additional constraints
on the product $Q_{N_5} Q_{D5}$ generalising \eqref{conchges}.
The Page charge would then change along the radial flow, reminiscent of the flows in \cite{Klebanov:2000hb}. 
Further exploration of the Page charges will be left for future work.

By following a similar calculation as in section \ref{ccc},
we find the central charge for the $AdS_3$ fixed point solutions is given by
\begin{align}
c=-\frac{3}{2} N Q_{N5} Q_{D5}(1-4Q^2)\,.
\end{align}
\subsection{T-duality and supersymmetry}
Starting with \eqref{fpfg} we introduce rescaled coordinates 
\begin{align}\label{pers2}
\bar z_1=\frac{2^{1/2}3^{3/2}(1-4Q^2)^{1/4}}{L^2} z_1\,,\qquad 
\bar z_2=\frac{3^{3/2}}{2^{1/2}L^2(1-4Q^2)^{1/4}} z_2\,,
\end{align}
and then carry out two T-dualities along the directions $\bar z_1,\bar z_2$ using, for example, 
the formulae in appendix B of \cite{Donos:2008ug}. Making the further change of ordinates
\begin{align}\label{coordchanges}
\alpha_1=\frac{1}{2}((1+2Q)\psi-\bar z_1)\,,\qquad 
\alpha_2=\frac{1}{2}((1-2Q)\psi+\bar z_1)\,,
\end{align}
we obtain
\begin{align}\label{tcsol}
ds^2&=\frac{2L^2(1-4Q^2)^{1/2}}{3^{3/2}}\left[ds^2(AdS_3)+\frac{2}{1+2Q} ds^2(S^3_1)+\frac{2}{1-2Q} ds^2(S^3_2)+d\bar z_2^2\right]\,,\nn
dC^2&=e^{-\phi_0}\frac{2L^2(1-4Q^2)^{1/2}}{3^{3/2}}\left[2 \mathrm{vol}(AdS_3) +\frac{4}{1+2Q} 
\mathrm{vol}(S^3_1)+\frac{4}{1-2Q} \mathrm{vol}(S^3_2)\right]\,,\nn
e^{\phi_0}&=\frac{3^{3/2}}{L^2}\,,
\end{align}
where, as before,
$ds^2(S^3_i)=\frac{1}{4}\left[ds^2_i+(d\alpha_i+P_i)^2\right]$. When $\alpha_i$ have period $4\pi$ this the general
$AdS_3\times S^3\times S^3\times S^1$ type IIB solution reviewed in appendix \ref{app:A}.

Observe that we again have $\partial_{\bar z_1}=\frac{1}{2}(-\partial_{\alpha_1}+\partial_{\alpha_2})$ and hence
following the same arguments as in section 2.5, we can conclude that the general
$AdS_3\times T^2\times S^2\times S^3$
solutions \eqref{fpfg} should preserve $(4,2)$ supersymmetry. This is confirmed in appendix \ref{appes}.

\section{Final comments}
We have constructed a novel class of type IIB supergravity solutions, preserving $(0,2$) supersymmetry, that interpolate between
$AdS_5\times T^{1,1}$ in the UV and a class of $AdS_3\times T^2\times S^2\times S^3$ solutions in the IR.
The IR solutions preserve $(4,2)$ superconformal supersymmetry and 
are related, locally, by two T-dualities to the well known $AdS_3\times S^3\times S^3\times S^1$ solutions. 
It would be interesting to establish in more detail how this T-duality works globally.
We examined the quantisation of Page charges for the $AdS_3\times T^2\times S^2\times S^3$ solutions, finding
some novel features. In particular, it does not seem possible to have 
the connection two-form $B$ of the gerbe be well defined as a two-form on an arbitrary five-manifold,
representing the homologically non-trivial five cycles, on which one wants to 
integrate in order to get the Page charge. However, for specific choices of the five-manifolds, we can find an appropriate connection, related by 
gauge transformations, so that
it is well defined. Furthermore, we found that a placing a constraint on 
the three-form fluxes ensured that the Page charges obtained in different gauges were all
integers. It would be helpful to investigate this in more detail as similar issues will arise in other supergravity solutions with fluxes.
One approach, in the present setting, is to try and make a precise connection with the globally realised T-duality.

It is known that classical type IIB string theory on $AdS_5\times T^{1,1}$ is not integrable \cite{Basu:2011di}. However, it seems likely that it will be integrable
on the $AdS_3\times T^2\times S^2\times S^3$ solutions we have discussed here (for related discussion see
\cite{Babichenko:2009dk,Sundin:2012gc,Wulff:2014kja}). 
It would be interesting to confirm this and also to investigate how the integrability emerges along the RG flow.

Another direction for further study would be to investigate whether the solutions that we have constructed here
can be generalised to solutions that flow from more general $AdS_5\times SE_5$ solutions, where $SE_5$ is a
five-dimensional Sasaki-Einstein solution. It is reasonable to expect that if we choose $SE_5$ to be one of the 
$Y^{p,q}$ spaces \cite{Gauntlett:2004yd} then there will be flows to the various 
$AdS_3\times T^2$ solutions found in section 4 of \cite{Donos:2008ug} (global aspects of the T-dual solutions are
discussed in \cite{Donos:2008hd}). 
These flow solutions might be difficult to construct explicitly, however, because unlike the case we have considered in this paper, there
is not a known consistent KK truncation on $Y^{p,q}$ analogous to the one on $T^{1,1}$.
These solutions would relate four dimensional superconformal field theories to two-dimensional
superconformal field theories with $(0,2)$ supersymmetry, complementing other such examples
\cite{Maldacena:2000mw,Cucu:2003bm,Cucu:2003yk,Almuhairi:2011ws,Donos:2011pn,Benini:2012cz,Benini:2013cda,Bobev:2014jva,Kutasov:2013ffl,Kutasov:2014hha}.

\section*{Acknowledgements}
We thank Stefano Cremonesi, John Estes, Chris Herzog, Juan Maldacena, Don Marolf, Eoin O Colgain, James Sparks, David Tong, Linus Wulff and 
especially Daniel Waldram for discussions. We thank the Aspen Center for Physics where some of this work was completed.
The work is supported in part by STFC grant ST/J0003533/1, EPSRC programme grant EP/K034456/1 and by the 
European Research Council under the European Union's Seventh Framework Programme (FP/2007-2013),
ERC Grant Agreements ERC-2013-AdG 339140 and STG 279943.

\appendix
\section{The $AdS_3\times S^3\times S^3\times S^1$ solution}\label{app:A}
Consider the standard $AdS_3\times S^3\times S^3\times S^1$ 
solution \cite{Cowdall:1998bu} (see also \cite{Gukov:2004ym}) which is supported by RR fluxes. In the conventions of section \ref{fluxq1} it can be written 
\begin{align}\label{rewrite}
ds^2&={\bar L}^2\left[ds^2(AdS_3)+r_1^2 ds^2(S^3_1)+r_2^2 ds^2(S^3_2)+dy^2\right]\,,\nn
dC^2&=\frac{1}{\bar g}\bar L^22\left[ \mathrm{vol}(AdS_3) +
r_1^2 \mathrm{vol}(S^3_1)+r_2^2 \mathrm{vol}(S^3_2)\right]\,,\nn
e^{\phi}&=1\,,
\end{align}
where $ds^2(S^3_i)$ are the standard round metrics on three-spheres, $y\cong y+\Delta y$ and,
in the notation of \eqref{tcsol},
\begin{align}
r_1^2=\frac{2}{1+2Q},\qquad r_2^2=\frac{2}{1-2Q}
\end{align}
with $0\le Q<1/2$. Observe that $r_1^2+r_2^2=r_1^2 r_2^2$.
We next quantise the flux. For the electric flux we have
\begin{align}
Q_{D1}=\frac{1}{(2\pi l_s)^6}\int_{S^3_1\times S^3_2\times S^1}*dC^2
=\left(\frac{\bar L}{l_s}\right)^6\frac{r_1^3r_2^3\Delta y}{8\bar g\pi^2}\in\mathbb{Z}
\end{align}
For the magnetic flux we have
\begin{align}
Q_{D5^{(1)}}=\frac{1}{(2\pi l_s)^2}\int_{S^3_1}dC^2=\left(\frac{\bar L}{l_s}\right)^2\frac{r_1^2}{\bar g}\in\mathbb{Z}\,,\nn
Q_{D5^{(2)}}=\frac{1}{(2\pi l_s)^2}\int_{S^3_2}dC^2=\left(\frac{\bar L}{l_s}\right)^2\frac{r_2^2}{\bar g}\in\mathbb{Z}\,.
\end{align}
Observe that we have
\begin{align}
4\pi \bar g Q_{D1}=\frac{\bar L\Delta y}{2\pi l_s}\bar g Q_{D5^{(1)}}\bar g Q_{D5^{(2)}}
\sqrt{\bar g Q_{D5^{(1)}}+\bar g Q_{D5^{(2)}}}\,,
\end{align}
which agrees with (2.17) of \cite{Gukov:2004ym} (they have set $2\pi l_s=1$), which shows that the radius of the circle is fixed by the RR charges $gQ$. Using \eqref{cchgeb},\eqref{actconv} we calculate the central charge
as \begin{align}\label{eq:ccharge1}
c=6Q_{D1}\frac{Q_{D5^{(1)}}Q_{D5^{(2)}}}{Q_{D5^{(1)}}+Q_{D5^{(2)}}}\,,
\end{align}
again agreeing with (2.20) of \cite{Gukov:2004ym}.

\section{Enhanced supersymmetry}\label{appes}
In this appendix we show that the $AdS_3\times\mathbb{R}^2\times S^2\times S^3$ solutions
given in \eqref{eq:AdS3_family} have a $(4,2)$ superconformal supersymmetry.
We set $L^2=3^{3/2}/(1-4Q^2)^{1/2}$ and use the orthonormal frame
\begin{align}
e^\mu&=\sqrt{2}\bar e^\mu,\quad \mu=0,1,4\,,\nn
e^2&=\frac{1}{(1-4Q^2)^{1/4}}dz_1,\qquad e^3 =\frac{1}{(1-4Q^2)^{1/4}}dz_2\,,\nn
e^5&=\frac{1}{(1+2Q)^{1/2}}d\theta_1,\quad e^6=\frac{1}{(1+2Q)^{1/2}}\sin\theta_1 d\phi_1,\nn
 e^7&=\frac{1}{(1-2Q)^{1/2}}d\theta_2,\quad e^8=\frac{1}{(1-2Q)^{1/2}}\sin\theta_2 d\phi_2\nn
e^9&=\frac{1}{\sqrt 2} (d\psi+P)\,,
\end{align}
where $\bar e^\mu$ is an orthonormal frame for a unit radius $AdS_3$. From
\eqref{eq:kill2} we deduce that
\begin{align}
(1-i\Gamma^{23})(1+\Gamma^{5678})\varepsilon=0\,.
\end{align}
We thus write
\begin{align}
&\varepsilon=\varepsilon_1+\varepsilon_2,
\end{align}
with 
\begin{align}
&\Gamma^{5678}\varepsilon_1=\varepsilon_1,\qquad \Gamma^{5678}\varepsilon_2=-\varepsilon_2,\nn
&\Gamma^{23}\varepsilon_1=-i\varepsilon_1\,.
\end{align}
The Killing spinor equations \eqref{eq:kill1} then take the following form. For the $\mu=0,1,4$ components we have
\begin{align}\label{kconsQ1}
&\bar\nabla_\mu (\varepsilon_1+\varepsilon_2)
-\frac{\left(1-4Q^{2} \right)^{\frac{1}{2}}}{4}\Gamma_\mu\Gamma^{256}\varepsilon_1^c
-\frac{i}{4}\Gamma_\mu\Gamma^{9}\left(\left(1+2Q\,\Gamma^{2356} \right)\varepsilon_1-(1-\Gamma^{2356})\varepsilon_2\right)=0, 
\end{align}
where $\bar\nabla\equiv\bar e^\mu\bar\nabla_\mu$ is the Levi-Civita connection on a unit radius $AdS_3$. 
For the 2,3 components we get
\begin{align}\label{23compsen}
&\left(1-4Q^{2} \right)^{\frac{1}{4}}\partial_{z_1} (\varepsilon_1+\varepsilon_2)
+\frac{\left(1-4Q^{2} \right)^{\frac{1}{2}}}{4\sqrt 2}\Gamma_2\Gamma^{256}\varepsilon_1^c\nn
&\qquad\qquad\qquad\qquad+\frac{i}{4\sqrt 2}\Gamma_2\Gamma^{9}\left(\left(-1+2Q\,\Gamma^{2356}\right)\varepsilon_1-(1+\Gamma^{2356})\varepsilon_2\right)=0\,,\nn
&\left(1-4Q^{2} \right)^{\frac{1}{4}}\partial_{z_2} (\varepsilon_1+\varepsilon_2)
+\frac{\left(1-4Q^{2} \right)^{\frac{1}{2}}}{4\sqrt 2}\Gamma_3\Gamma^{256}\varepsilon_1^c\nn
&\qquad\qquad\qquad\qquad+\frac{i}{4\sqrt 2}\Gamma_3\Gamma^{9}\left(\left(-1+2Q\,\Gamma^{2356}\right)\varepsilon_1-(1+\Gamma^{2356})\varepsilon_2\right)=0\,.
\end{align}
For the 5,6 components we get
\begin{align}
&\left(\left(1+2Q\right)^{\frac{1}{2}}\,\partial_{\theta_1}+\frac{1+2Q}{4\sqrt 2}\Gamma^{69}\right) (\varepsilon_1+\varepsilon_2)
+\frac{\left(1-4Q^{2} \right)^{\frac{1}{2}}}{4\sqrt 2}\Gamma_5\Gamma^{256}\left(\varepsilon_1^c+(1-i\Gamma^{23})\varepsilon_2^c\right)\nn
&\qquad\qquad\qquad\qquad\qquad\qquad\quad +\frac{i}{4\sqrt 2}\Gamma_5\Gamma^{9}\left((1+\Gamma^{2356})\varepsilon_1-\left(1-2Q\,\Gamma^{2356} \right)\varepsilon_2\right)=0\,,\nn
&\left(\frac{\left(1+2Q\right)^{\frac{1}{2}}}{\sin\theta_1}\partial_{\phi_1}+\frac{\left(1+2Q\right)^{\frac{1}{2}}}{2}\cot\theta_1(2\partial_\psi-\Gamma^{56})-\frac{1+2Q}{4\sqrt 2}\Gamma^{59}\right) (\varepsilon_1+\varepsilon_2)\nn
&
+
\frac{\left(1-4Q^{2} \right)^{\frac{1}{2}}}{4\sqrt 2}\Gamma_6\Gamma^{256}\left(\varepsilon_1^c+(1-i\Gamma^{23})\varepsilon_2^c\right)
+\frac{i}{4\sqrt 2}\Gamma_6\Gamma^{9}\left((1+\Gamma^{2356})\varepsilon_1-\left(1-2Q\,\Gamma^{2356} \right)\varepsilon_2\right)=0\,.
\end{align}
For the 7,8 components we get
\begin{align}
&\left(\left(1-2Q\right)^{\frac{1}{2}}\partial_{\theta_2}+\frac{1-2Q}{4\sqrt 2}\Gamma^{89}\right) (\varepsilon_1+\varepsilon_2)
+\frac{\left(1-4Q^{2} \right)^{\frac{1}{2}}}{4\sqrt 2}\Gamma_7\Gamma^{256}\left(\varepsilon_1^c
-(1-i\Gamma^{23})\varepsilon_2^c\right)\nn
&\qquad\qquad\qquad\qquad\qquad\qquad\quad+\frac{i}{4\sqrt 2}\Gamma_7\Gamma^{9}\left(
(1-\Gamma^{2356})\varepsilon_1-\left(1+2Q\,\Gamma^{2356} \right)\varepsilon_2\right)=0\,,\nn
&\left(\frac{\left(1-4Q^{2} \right)^{\frac{1}{2}}}{\sin\theta_2}\partial_{\phi_2}+\frac{\left(1-4Q^{2} \right)^{\frac{1}{2}}}{2}\cot\theta_2(2\partial_\psi-\Gamma^{78})-\frac{1-2Q}{4\sqrt 2}\Gamma^{79}\right) (\varepsilon_1+\varepsilon_2)\nn&
+
\frac{\left(1-4Q^{2} \right)^{\frac{1}{2}}}{4\sqrt 2}\Gamma_8\Gamma^{256}\left(\varepsilon_1^c-(1-i\Gamma^{23})\varepsilon_2^c\right)
+\frac{i}{4\sqrt 2}\Gamma_8\Gamma^{9}\left(
(1-\Gamma^{2356})\varepsilon_1-\left(1+2Q\,\Gamma^{2356} \right)\varepsilon_2\right)=0\,.
\end{align}
Finally, for the 9 component we get
\begin{align}\label{9compsen}
&\partial_\psi(\varepsilon_1+\varepsilon_2)-\frac{1}{4}\Gamma^{56}\varepsilon_2-\frac{Q}{{2}}\,\Gamma^{56}\varepsilon_{1}
-\frac{\left(1-4Q^{2} \right)^{\frac{1}{2}}}{8}\Gamma_9\Gamma^{256}\varepsilon_1^c\nn
&\qquad\qquad\qquad\qquad\qquad\qquad\qquad+\frac{i}{8}\left(\left(1+2Q\,\Gamma^{2356} \right)\varepsilon_1-(1-\Gamma^{2356})\varepsilon_2\right)=0\,.
\end{align}

By examining the integrability conditions for the two equations involving derivatives with {appes}ct to $z_1,z_2$, \eqref{23compsen}, 
we deduce that
\begin{align}
\Gamma^{2569}\varepsilon_1=\frac{i}{\left(1-4Q^{2}\right)^{\frac{1}{2}}}\,\left(1-2i Q\,\Gamma^{56} \right)\,\varepsilon_1^c
\end{align}
and also that $\partial_{z_i}\varepsilon_1=0$.
Examining \eqref{9compsen} we also see that
$\partial_\psi \varepsilon_1=0$.
From \eqref{kconsQ1} we can deduce that
\begin{align}\label{bhl}
&\bar\nabla_\mu \varepsilon_1
-\frac{i}{ 2}\Gamma_\mu\Gamma^{9}\varepsilon_1=0,\nn
&\bar\nabla_\mu \varepsilon_2
+\frac{i}{4}\Gamma_\mu\Gamma^{9}\left(1-\Gamma^{2356}\right)\varepsilon_2=0, 
\end{align}
and the integrability conditions for the second line implies
\begin{align}
\Gamma^{2356}\varepsilon_2=-\varepsilon_2\,.
\end{align}
As a consequence we conclude from \eqref{23compsen} that 
$\partial_{z_i}\varepsilon_2=0$.

It is now convenient to further decompose
\begin{align}
\varepsilon_{2}=\varepsilon_{2}^{+}+\varepsilon_{2}^{-},\qquad\qquad
i \Gamma^{23}\varepsilon_{2}^{\pm}=\pm\,\varepsilon_{2}^{\pm}\,.
\end{align}
After projecting out the equations using $(1\pm i\Gamma^{23})/2$ and $(1\pm i\Gamma^{56})/2$ we deduce the following general
solution. Firstly, $\varepsilon_1$ and  $\varepsilon^+_2$ are given by
\begin{align}
\varepsilon_{1}=&\cos\left(\frac{\theta_{1}}{2} \right)\,\cos\left(\frac{\theta_{2}}{2}\right)\,\left[({1-2Q})^{1/2}c_{1}-(1+2Q)^{1/2}\,\Gamma^{29}c_{1}^{c}\right]\nn
+&\cos\left(\frac{\theta_{1}}{2} \right)\,\sin\left(\frac{\theta_{2}}{2}\right)\,\left[(1-2Q)^{1/2}c_{2}-(1+2Q)^{1/2}\,\Gamma^{29}c_{2}^{c} \right]\nn
+&\sin\left(\frac{\theta_{1}}{2} \right)\,\cos\left(\frac{\theta_{2}}{2}\right)\,\Gamma^{68}\left[(1+2Q)^{1/2}\,c_{2}+(1-2Q)^{1/2}\Gamma^{29}c_{2}^{c} \right]\nn
-&\sin\left(\frac{\theta_{1}}{2} \right)\,\sin\left(\frac{\theta_{2}}{2}\right)\,\Gamma^{68}\left[(1+2Q)^{1/2}\,c_{1}+(1-2Q)^{1/2}\Gamma^{29}c_{1}^{c} \right]
\end{align}
and
\begin{align}
\varepsilon_{2}^{+}=&\sqrt{2}\Bigg[-\,\cos\left(\frac{\theta_{1}}{2} \right)\,\sin\left(\frac{\theta_{2}}{2}\right)\,\Gamma^{89}\,c_{1}
+\,\cos\left(\frac{\theta_{1}}{2} \right)\,\cos\left(\frac{\theta_{2}}{2}\right)\,\Gamma^{89}c_{2}\nn
&\qquad +\,\sin\left(\frac{\theta_{1}}{2} \right)\,\sin\left(\frac{\theta_{2}}{2}\right)\,\Gamma^{26}c_{2}^{c}
+\,\sin\left(\frac{\theta_{1}}{2} \right)\,\cos\left(\frac{\theta_{2}}{2}\right)\,\Gamma^{26}c_{1}^{c}\Bigg]\,,
\end{align}
with
\begin{align}
c_{1}=e^{-\frac{i}{2}\left(\phi_{1}-\phi_{2} \right)}\,d_{1},\quad c_{2}=e^{-\frac{i}{2}\left(\phi_{1}+\phi_{2}\right)}\,d_{2}\,,
\end{align}
and the two ten-dimensional spinors $d_{a}$, $a=1,2$, satisfy the projections
\begin{align}\label{genqprojsi}
i\,\Gamma^{23}d_{a}=d_{a},\quad i\Gamma^{56}d_{a}=d_{a},\quad i\Gamma^{78}d_{a}=-d_{a},\qquad a=1,2\,.
\end{align}
The $d_a$ only depend on the coordinates on $AdS_{3}$ and should satisfy
\begin{align}\label{diadseqf}
(\bar\nabla_\mu 
-\frac{i}{ 2}\Gamma_\mu\Gamma^{9})d_{a}=0,\qquad \qquad a=1,2\,.
\end{align}

Secondly, the parameter $\varepsilon_{2}^{-}$ takes the simple form
\begin{align}
\varepsilon_{2}^{-}=e^{i\frac{\psi}{2}}\,d_{3}\,,
\end{align}
with 
\begin{align}\label{genqprojs3}
i\,\Gamma^{23}d_{3}=d_{3},\quad i\Gamma^{56}d_{3}=d_{3},\quad i\Gamma^{78}d_{3}=d_{3}\,.
\end{align}
Again $d_3$ only depends on the coordinates on $AdS_{3}$ and now should satisfy
\begin{align}\label{d3adseqf}
(\bar\nabla_\mu 
+\frac{i}{ 2}\Gamma_\mu\Gamma^{9})d_{3}=0\,.
\end{align}

To solve \eqref{diadseqf}, \eqref{d3adseqf} we use the following frame and coordinates for
$AdS_3$:
\begin{align}
\bar e^0=e^{\rho} dt,\qquad \bar e^1=e^{\rho} dx,\qquad \bar e^4=d\rho\,,
\end{align}
to obtain
\begin{align}
d_1&=e^{\rho/2}\alpha_1^-+\left[e^{-\rho/2}-e^{\rho/2}(t\Gamma_0+x\Gamma_1)\Gamma_{4}\right]\alpha_1^+\,,\nn
d_2&=e^{\rho/2}\alpha_2^-+\left[e^{-\rho/2}-e^{\rho/2}(t\Gamma_0+x\Gamma_1)\Gamma_{4}\right]\alpha_2^+\,,\nn
d_3&=e^{\rho/2}\alpha_3^++\left[e^{-\rho/2}-e^{\rho/2}(t\Gamma_0+x\Gamma_1)\Gamma_{4}\right]\alpha_3^-\,,
\end{align}
where $\alpha_1^\pm,\alpha_2^\pm$ and $\alpha_3^\pm$  satisfy the projections
\eqref{genqprojsi} and \eqref{genqprojs3}, respectively, and in addition
\begin{align}
\Gamma^{01}\alpha_a^\pm=\pm\alpha_a^\pm,\quad a=1,2,\qquad
\Gamma^{01}\alpha_3^\pm=\pm\alpha_3^\pm\,.
\end{align}
We see that $\alpha_3^+$ parametrises the $(0,2)$ Poincar\'e supersymmetry that is preserved throughout the whole
flow of the domain wall solutions (recall \eqref{eq:projections} and \eqref{proj2}).
The $\alpha_a^-$ parametrise an enhancement of the Poincar\'e supersymmetry to $(4,2)$.
The remaining six supersymmetries, labelled by $\alpha_3^-$ and $\alpha_a^+$, parametrise the
superconformal supersymmetries.

As noted in the text, using the results of \cite{Gauntlett:1998kc} we can deduce that the superisometry algebra is of the form $ D(2,1|\alpha)\times G$, where $G\subset D(2,1|\alpha)$ and has a bosonic sub-algebra given
by $SL(2,\mathbb{R})\times U(1)^2$. This could be verified using the explicit Killing spinors that we have constructed, but we shall not do that here.

\section{Page charge quantisation}\label{pchge}
We will discuss the essential aspects of the quantisation of Page charges that we employed in
the bulk of the paper in a simplified setting.
We suppose that we have a manifold with topology $S^2\times S^3$ which is presented as a circle bundle fibred over $S^2_1\times S^2_2$ exactly 
as for $T^{1,1}$:
\begin{align}
ds^2(S^2\times S^3)=c_1^2 ds^2_1+c_2^2 ds^2_2+c_3^2 D\psi^2\,,
\end{align}
where $c_i$ are non-zero constants which won't be important, and
\begin{align}
ds^2_i&=d\theta^2_i+\sin^2\theta_id\phi_i^2\,,\nn
d(D\psi)&=vol_1+vol_2\,,\nn
\vol_i&=\sin\theta_i d\theta_i \wedge d\phi_i,\qquad \text{no sum on $i$}\,,
\end{align}
and $\psi$ has period $4\pi$. We note that $D\psi$ is a globally defined one-form.
A positive orientation on $S^2\times S^3$ is given by $D\psi\wedge \mathrm{vol}_1\wedge \mathrm{vol}_2$.

{\bf Topology:}
A smooth manifold $S^3_1$ that can be used to generate $H_3(S^2\times S^3,\mathbb{Z})$
is provided
by the circle bundle restricted to the $S^2_1$ factor on the base space.  We can also choose the 
circle bundle restricted to the $S^2_2$ factor on the base space, with {\it opposite orientation}, which we call $S^3_2$. 
Observe that
$D\psi\wedge (\mathrm{vol}_1-\mathrm{vol}_2)$ is closed (since
 $d(D\psi)=\mathrm{vol}_1+
\mathrm{vol}_2$) and hence when it is integrated over a three-cycle, it will only depend
on the homology class of the cycle. We have:
\begin{align}\label{bipone}
\int_{[S^3]} D\psi\wedge (\mathrm{vol}_1-\mathrm{vol}_2)=\int_{S^3_1} D\psi\wedge (\mathrm{vol}_1)=\int_{S^3_2} D\psi\wedge (-\mathrm{vol}_2)=16\pi^2\,.
\end{align}

To find a smooth manifold that can be used to generate $H_2(S^2\times S^3,\mathbb{Z})$ we consider
any smooth manifold $S$ on the base that represents the cycle $[S]=[S^2_2]-[S^2_1]$. Since the circle bundle is trivial over $S$, there is a section $s$ and we can uses $s(S)$ to generate $H_2(S^2\times S^3,\mathbb{Z})$. We can take, for example,
$\theta_1=\theta_2$ and $-\phi_1=\phi_2$ at fixed $\psi$. In particular we have
\begin{align}\label{bip}
-\int_{s(S)} \mathrm{vol}_1=\int_{s(S)} \mathrm{vol}_2=4\pi\,.
\end{align}

It can be helpful to explicitly identify the Poincar\'e duals of the above generators.
A representative closed 3-form generator, $\Phi_3$, of $H^3(S^2\times S^3,\mathbb{Z})$ is given by $\Phi_3=(1/16\pi^2)D\psi\wedge (\mathrm{vol}_1-\mathrm{vol}_2)$ with the property that $\int_E \Phi_3=1$, where $E$ generates $H_3(S^2\times S^3,\mathbb{Z})$. 
The three-form $\Phi_3$ is Poincar\'e dual to $[S]$ and we can use it to evaluate
$\int_{s(S)}\omega_2=\int_{S^2\times S^3}\omega_2\wedge \Phi_3$ for any closed two-form $\omega_2$. In particular, we can check \eqref{bip}.
A representative closed 2-form generator, $\Phi_2$, of $H^2(S^2\times S^3,\mathbb{Z})$ 
is given by $\Phi_2=-\frac{1}{8\pi}(\mathrm{vol}_1-\mathrm{vol}_2)$ with the property that $\int_{s(S)} \Phi_2=1$. 
The two-form $\Phi_2$ is Poincar\'e dual to $[S^3]$ and we can use it to evaluate
$\int_{[S^3]}\omega_3=\int_{S^2\times S^3}\omega_3 \wedge \Phi_2$ for any closed three-form $\omega_3$. 
In particular, we can check \eqref{bipone}. Note that $\int_{S^2\times S^3} \Phi_2\wedge\Phi_3=1$.

{\bf Patches:}
Let us introduce four coordinate patches to cover $S^2\times S^3$.
We consider four patches $U_{NN},U_{NS},U_{SN},U_{SS}$, isomorphic to $\mathbb{R}^4\times S^1$. We take $U_{NN}$ 
to consist of the northern hemispheres of the two $S^2$'s on the base as well as a coordinate $\psi_{NN}$ with period $4\pi$.
Next, $U_{NS}$ is the northern hemisphere of $S^2_1$ and the southern hemisphere of $S^2_2$ on the base, 
as well as a coordinate $\psi_{NS}$ with period $4\pi$, and similarly for the rest. Now we know that the one-form 
$D\psi\equiv d\psi+P$ is globally defined and we have
\begin{align}
D\psi
&=d\psi_{NN}+(1-\cos\theta_1)d\phi_1+(1-\cos\theta_2)d\phi_2\,,\nn
&=d\psi_{NS}+(1-\cos\theta_1)d\phi_1+(-1-\cos\theta_2)d\phi_2\,,\nn
&=d\psi_{SN}+(-1-\cos\theta_1)d\phi_1+(1-\cos\theta_2)d\phi_2\,,\nn
&=d\psi_{SS}+(-1-\cos\theta_1)d\phi_1+(-1-\cos\theta_2)d\phi_2\,.
\end{align}
On the overlaps of the patches we have
\begin{align}\label{patchc}
\psi_{NN}=\psi_{NS}-2\phi_2=\psi_{SN}-2\phi_1=\psi_{SS}-2\phi_1-2\phi_2\,,
\end{align}
which shows that we have a good circle bundle, with the patching done with $U(1)$ gauge-transformations: e.g. $\psi_{NN}/2=\psi_{NS}/2=ie^{-i\phi_2}d(e^{i\phi_2})$ (the factors of two here are because $\psi$ has period $4\pi$).

{\bf Fluxes and charges:}
To illustrate the main features of the calculation in the text, we will consider a slightly simpler problem where
we ``forget" the $T^2$ factor. We could imagine that we have carried out a dimensional reduction on the $T^2$,
for example. The advantage of doing this is that the ambiguities in defining Page charges will just involve
gauge-transformations of $U(1)$ gauge-connections rather than gerbes.

We consider, therefore, the following globally defined fluxes
\begin{align}
F_3&=\frac{kl}{16}D\psi\wedge (vol_1+vol_2)\,,\nn
F_2&=-\frac{k}{4}(vol_1-vol_2)\,,\nn
G_2&=\frac{l}{4}(vol_1-vol_2)\,,
\end{align}
with $dF_2=dG_2=0$ and $dF_3=F_2\wedge G_2$. We will assume that $F_2$ and $G_2$ are the curvature two-forms for two $U(1)$ connections with integer Chern numbers. Thus we demand that 
$k,l\in\mathbb{Z}$ so that that we have the quantisation conditions:
\begin{align}
\frac{1}{2\pi}\int_{s(S)}F_2=k,\qquad \frac{1}{2\pi}\int_{s(S)}G_2=-l\,.
\end{align}

If we write $F_2=dA_1$, a natural Page charge to consider quantising is
\begin{align}
\frac{1}{(2\pi)^2}\int_{[S^3]} (F_3-A_1\wedge G_2)\,.
\end{align}
For definiteness we define $S^3_1$ and $S^3_2$ to sit at a fixed point on the northern hemisphere of the other two-sphere.
We can calculate
\begin{align}
\frac{1}{(2\pi)^2}\int_{S^3_1} F_3 =+\frac{kl}{4},\qquad \frac{1}{(2\pi)^2}\int_{S^3_2} F_3 =-\frac{kl}{4}\,,
\end{align}
which differ because $F_3$ is not closed.
We now introduce two gauge connections given by
\begin{align}
A^{(1)}_1&=-\frac{k}{4}(d\psi_{NN}+(1-\cos\theta_1)d\phi_1 - (1-\cos\theta_2)d\phi_2)\,,\nn
 A^{(2)}_1&=-\frac{k}{4}(-d\psi_{NN}+(1-\cos\theta_1)d\phi_1-(1-\cos\theta_2) d\phi_2)\,.
\end{align}
Being connections with cohomologically non-trivial field strengths, these cannot be globally defined one-forms.
However, they should patch together using $U(1)$ gauge transformations.
Let us first consider $A^{(1)}_1$. It is clearly defined on the $NN$ patch. It is also well defined on the $SN$ patch 
after using \eqref{patchc}. Next, moving to the $NS$ coordinate patch we get
something that is well defined up to a $U(1)$ gauge transformation:
\begin{align}
A^{(1)}_1&= -\frac{k}{4}(d\psi_{NS}+(1-\cos\theta_1)d\phi_1+(1+\cos\theta_2)d\phi_2) +kd\phi_2\,,\nn
&= -\frac{k}{4}(d\psi_{NS}+(1-\cos\theta_1)d\phi_1+(1+\cos\theta_2)d\phi_2) -ie^{-ik\phi_2}d(e^{ik\phi_2})\,,
\end{align}
where we recall that $\phi_2$ has period $2\pi$. Moving to $SS$ is similar. Thus $A^{(1)}_1$ is a $U(1)$ gauge connection for $F_2$.
Furthermore, we observe that $A^{(1)}_1$ is a globally well defined one-form on $S^3_1$ 
since for a fixed point on the northern hemisphere patch of the $S^2_2$
we can switch to the $SN$ patch in a regular manner using \eqref{patchc}.

Similar comments apply to $A^{(2)}_1$.
We calculate
\begin{align}
A^{(1)}_1-A^{(2)}_1=-\frac{k}{2}d\psi_{NN}=-ie^{-ik\psi/2}d(e^{ik\psi/2})\,,
\end{align}
which shows that they are related by a good $U(1)$ gauge transformation, since $\psi$ has period $4\pi$.
Thus $A^{(2)}_1$ is also $U(1)$ gauge connection for $F_2$ and, in contrast to $A^{(1)}_1$, is now a
well defined one-form on $S^3_2$.

We can now calculate:
\begin{align}
\frac{1}{(2\pi)^2}\int_{S^3_1} -A^{(1)}_1\wedge G_2=+\frac{kl}{4},\qquad \frac{1}{(2\pi)^2}\int_{S^3_2} -A^{(2)}\wedge G_2=-\frac{kl}{4}\,,
\end{align}
and hence
\begin{align}
\frac{1}{(2\pi)^2}\int_{S^3_1} (F_3-A^{(1)}_1\wedge G_2)=\frac{kl}{2},\qquad
\frac{1}{(2\pi)^2}\int_{S^3_2} (F_3-A^{(2)}_1\wedge G_2)=-\frac{kl}{2}\,.
\end{align}
Now $F_3-A_1\wedge G_2$ is closed, so we may naively have thought that these should be equal.
However, $A_1$ is connection, so $F_3-A_1\wedge G_2$ is not a three-form and does not define a cohomology class.
If we demand that $kl=2\bar N$ with $\bar N\in\mathbb{Z}$ then both of these are integers.

In essence this is the flux quantisation procedure that we have adopted in the main text. An open issue, which we leave for the future,
is to determine what happens if we choose other smooth three-manifolds $\Sigma$ to represent $H_3$.
What are the conditions for
there to exist a connection one-form, related to $A^{(i)}$ by a gauge transformation,
which is well defined on $\Sigma$ and, when it does exist, is the corresponding Page charge always an integer times $\bar N$?
We believe that similar issues will arise in other contexts.

\bibliographystyle{utphys}
\bibliography{helical}{}
\end{document}